\documentclass[ALICE,manyauthors]{cernphprep}
\usepackage[comma,square,numbers,sort&compress]{natbib}
\usepackage{graphicx}
\usepackage{lineno}
\usepackage{color}
\usepackage{hyperref}
\usepackage{caption}
\usepackage{csquotes}
\usepackage[Symbol]{upgreek}
\extrafloats{100}

\usepackage[T1]{fontenc}
\usepackage{orcidlink}

\begin{document}
\newcommand {\red}[1]    {\textcolor{red}{#1}}
\newcommand {\green}[1]  {\textcolor{green}{#1}}
\newcommand {\blue}[1]   {\textcolor{blue}{#1}}
\newcommand {\dgreen}[1]    {\textcolor{darkgreen}{#1}}
\newcommand {\orange}[1]   {\textcolor{orange}{#1}}

\newcommand{\deutsch}[1]{\foreignlanguage{german}{#1}}

\newcommand{\note}[1]{\blue{{\textbf{NOTE:} #1}}}
\newenvironment{oldconvert}
{\textbf{POTENTIALLY CONVERT/UPDATE/USE:}\itshape\color{gray}}

{\color{black}}
\newcommand{\TBC}       {\textcolor{red}{\footnotesize \textsc{TBC}}}

\DeclareRobustCommand{\unit}[2][]{%
        \begingroup%
                \def\0{#1}%
                \expandafter%
        \endgroup%
        \ifx\0\@empty%
                \ensuremath{\mathrm{#2}}%
        \else%
                \ensuremath{#1\,\mathrm{#2}}%
        \fi%
        }
\DeclareRobustCommand{\unitfrac}[3][]{%
        \begingroup%
                \def\0{#1}%
                \expandafter%
        \endgroup%
        \ifx\0\@empty%
                \raisebox{0.98ex}{\ensuremath{\mathrm{\scriptstyle#2}}}%
                \nobreak\hspace{-0.15em}\ensuremath{/}\nobreak\hspace{-0.12em}%
                \raisebox{-0.58ex}{\ensuremath{\mathrm{\scriptstyle#3}}}%
        \else
                \ensuremath{#1}\,%
                \raisebox{0.98ex}{\ensuremath{\mathrm{\scriptstyle#2}}}%
                \nobreak\hspace{-0.15em}\ensuremath{/}\nobreak\hspace{-0.12em}%
                \raisebox{-0.58ex}{\ensuremath{\mathrm{\scriptstyle#3}}}%
        \fi%
}

%
%
\newcommand{\ie}{i.\,e.\;}
\newcommand{\eg}{e.\,g.\;}

\newcommand{\run}[1]{\textsc{Run\,#1}}

%
%

\newcommand{\dNdeta}{\ensuremath{\mathrm{d}N_\mathrm{ch}/\mathrm{d}\eta}\xspace}

%
%
\newlength{\smallerpicsize}
\setlength{\smallerpicsize}{70mm}
\newlength{\smallpicsize}
\setlength{\smallpicsize}{90mm}
\newlength{\mediumpicsize}
\setlength{\mediumpicsize}{120mm}
\newlength{\largepicsize}
\setlength{\largepicsize}{150mm}

\newcommand{\PICX}[6]{
   \begin{figure}[#6]
      \begin{center}
         \vspace{3ex}
         \includegraphics[width=#3]{#1}
         \caption[#4]{\label{#2} #5}
      \end{center}
   \end{figure}
}

\newcommand{\DPICX}[7]{
   \begin{figure}[#7]
      \begin{center}
         \vspace{3ex}
         \includegraphics[width=#4]{#1}
         \includegraphics[width=#4]{#2}
         \caption[#5]{\label{#3} #6}
      \end{center}
   \end{figure}
}

\newcommand{\PICH}[5]{
   \begin{figure}[H]
      \begin{center}
         \vspace{3ex}
         \includegraphics[width=#3]{#1}
         \caption[#4]{\label{#2} #5}
      \end{center}
   \end{figure}
}

%
%
%
\newcommand{\figs}{Figs.\xspace}
\newcommand{\Figs}{Figures\xspace}
\newcommand{\eqn}{equation\xspace}
\newcommand{\Eqn}{Equation\xspace}
\newcommand{\figref}[1]{Fig.~\ref{#1}}
\newcommand{\figsref}[2]{Figs.~\ref{#1}--~\ref{#2}}
\newcommand{\Figref}[1]{Figure~\ref{#1}}
\newcommand{\tabref}[1]{Tab.~\ref{#1}}
\newcommand{\Tabref}[1]{Table~\ref{#1}}
\newcommand{\appref}[1]{App.~\ref{#1}}
\newcommand{\Appref}[1]{Appendix~\ref{#1}}
\newcommand{\secs}{Secs.\xspace}
\newcommand{\Secs}{Sections\xspace}
\newcommand{\secref}[1]{Sec.~\ref{#1}}
\newcommand{\Secref}[1]{Section~\ref{#1}}
\newcommand{\chaps}{Chaps.\xspace}
\newcommand{\Chaps}{Chapters\xspace}
\newcommand{\chapref}[1]{Chap.~\ref{#1}}
\newcommand{\Chapref}[1]{Chapter~\ref{#1}}
\newcommand{\lstref}[1]{Listing~\ref{#1}}
\newcommand{\Lstref}[1]{Listing~\ref{#1}}
%
%
\newcommand{\otoprule}{\midrule[\heavyrulewidth]}
\topfigrule
%
\newcommand {\stat}     {({\it stat.})~}
\newcommand {\syst}     {({\it syst.})~}
 \newcommand {\mom}       {\ensuremath{p}}
\newcommand {\pT}        {\pt}
\newcommand {\meanpT}    {\ensuremath{\langle p_{\mathrm{T}} \kern-0.1em\rangle}\xspace}
\newcommand {\mean}[1]   {\ensuremath{\langle #1 \kern-0.1em\rangle}\xspace}
\newcommand {\sqrtsNN}   {\ensuremath{\sqrt{s_{\textsc{NN}}}}\xspace}
\newcommand {\sqrts}     {\ensuremath{\sqrt{s}}\xspace}
\newcommand {\vf}        {\ensuremath{v_{\mathrm{2}}}\xspace}
\newcommand {\et}        {\ensuremath{E_{\mathrm{t}}}\xspace}
\newcommand {\mT}        {\ensuremath{m_{\mathrm{T}}}\xspace}
\newcommand {\mTmZero}   {\ensuremath{m_{\mathrm{T}} - m_0}\xspace}
\newcommand {\minv}      {\mbox{$m_{\ee}$}}
\newcommand {\rap}       {\mbox{$y$}}
\newcommand {\absrap}    {\mbox{$\left | y \right | $}}
\newcommand {\rapXi}     {\mbox{$\left | y(\rmXi) \right | $}}
\newcommand {\abspseudorap} {\mbox{$\left | \eta \right | $}}
\newcommand {\pseudorap} {\mbox{$\eta$}}
\newcommand {\cTau}      {\ensuremath{c\tau}}
\newcommand {\sigee}     {$\sigma_E$/$E$}
\newcommand {\dNdy}      {\ensuremath{\mathrm{d}N/\mathrm{d}y}}
\newcommand {\dNdpt}     {\ensuremath{\mathrm{d}N/\mathrm{d}\pT }}
\newcommand {\dNdptdy}   {\ensuremath{\mathrm{d^{2}}N/\mathrm{d}\pT\mathrm{d}y }}
\newcommand {\fracdNdptdy}   {\ensuremath{ \frac{\mathrm{d^{2}}N}{\mathrm{d}\pT\mathrm{d}y } }}
\newcommand {\dNdmtdy}   {\ensuremath{\mathrm{d^{2}}N/\mathrm{d}\mT\mathrm{d}y }}
\newcommand {\dN}        {\ensuremath{\mathrm{d}N }}
\newcommand {\dNsquared} {\ensuremath{\mathrm{d^{2}}N }}
\newcommand {\dpt}       {\ensuremath{\mathrm{d}\pT }}
\newcommand {\dy}        {\ensuremath{\mathrm{d}y}}
\newcommand {\dNdyBold}  {\ensuremath{\boldsymbol{\dN/\dy}}\xspace}
\newcommand {\dNchdy}    {\ensuremath{\mathrm{d}N_\mathrm{ch}/\mathrm{d}y }\xspace}
\newcommand {\dNchdeta}  {\ensuremath{\mathrm{d}N_\mathrm{ch}/\mathrm{d}\eta }\xspace}
\newcommand {\dNchdptdeta}  {\ensuremath{\mathrm{d}N_\mathrm{ch}/\mathrm{d}\pT\mathrm{d}\eta }\xspace}
\newcommand {\Raa}       {\ensuremath{R_\mathrm{AA}}}
\newcommand {\RpPb}       {\ensuremath{R_\mathrm{pPb}}\xspace}
\newcommand {\Nevt}      {\ensuremath{N_\mathrm{evt}}}
\newcommand {\NevtINEL}  {\ensuremath{N_\mathrm{evt}(\textsc{inel})}}
\newcommand {\NevtNSD}   {\ensuremath{N_\mathrm{evt}(\textsc{nsd})}}
\newcommand{\dEdx}       {\ensuremath{\mathrm{d}E/\mathrm{d}x}\xspace}
\newcommand{\ttof}       {\ensuremath{t_\mathrm{TOF}}\xspace}
\newcommand {\ee}        {\mbox{$\mathrm {e^+e^-}$}\xspace}
\newcommand {\ep}        {\mbox{$\mathrm {e\kern-0.05em p}$}\xspace}
\newcommand {\pp}        {\mbox{$\mathrm {p\kern-0.05em p}$}\xspace}
\newcommand {\pplong}        {\mbox{\textit{proton--proton}}\xspace}
\newcommand {\ppBoldMath} {\mbox{$\mathrm { \mathbf p\kern-0.05em \mathbf p }$}\xspace}
\newcommand {\ppbar}     {\mbox{$\mathrm {p\overline{p}}$}\xspace}
\newcommand {\PbPb}      {\ensuremath{\mbox{Pb--Pb}}\xspace}
\newcommand {\AuAu}      {\ensuremath{\mbox{Au--Au}}\xspace}
\newcommand {\CuCu}      {\ensuremath{\mbox{Cu--Cu}}\xspace}
\renewcommand {\AA}      {\ensuremath{\mbox{A--A}}\xspace}
\newcommand {\pPb}       {\ensuremath{\mbox{p--Pb}}\xspace}
\newcommand {\Pbp}       {\ensuremath{\mbox{Pb--p}}\xspace}
\newcommand {\hPM}       {\ensuremath{h^{\pm}}\xspace}
\newcommand {\rphi}      {\ensuremath{(r,\phi)}\xspace}
\newcommand {\alphaS}    {\ensuremath{ \alpha_s}\xspace}
\newcommand {\MeanNpart} {\mbox{\ensuremath{< \kern-0.15em N_{part} \kern-0.15em >}}}

\newcommand {\sig}       {\ensuremath{S}\xspace}
\newcommand {\expsig}    {\ensuremath{\hat{S}}\xspace}
\newcommand {\prob}      {\ensuremath{P}\xspace}
\newcommand {\prior}     {\ensuremath{C}\xspace}
\newcommand {\prop}      {\ensuremath{F}\xspace}
\newcommand {\atrue}     {\ensuremath{\vec{A}_{\mathrm{true}}}\xspace}
\newcommand {\ameas}     {\ensuremath{\vec{A}_{\mathrm{meas}}}\xspace}
\newcommand {\detresp}   {\ensuremath{R}\xspace}

\newcommand {\pid}       {\ensuremath{\mathrm{\epsilon}_\mathrm{PID}}\xspace}
\newcommand {\nsigma}    {\ensuremath{\mathrm{n_{\sigma}}}\xspace}
\newcommand {\ylab} {\ensuremath{\mathrm{| y_{lab} |}}\xspace}

\newcommand {\vzeroperc}        {\mbox{$\mathrm {V0M}_\mathrm{perc}$}\xspace}
\newcommand {\ntrkl}        {\mbox{$n_\mathrm{trkl}$}\xspace}

%
%
\newcommand {\mass}         {\mbox\mathrm{MeV$\kern-0.15em /\kern-0.12em c^2$}}
\newcommand {\tev}          {\ensuremath{\mathrm{TeV}}\xspace}
\newcommand {\gev}          {\ensuremath{\mathrm{GeV}}\xspace}
\newcommand {\mev}          {\ensuremath{\mathrm{MeV}}\xspace}
\newcommand {\kev}          {\ensuremath{\mathrm{keV}}\xspace}
\newcommand {\tevBoldMath}  {\ensuremath{\mathrm{\pmb{TeV}}}}
\newcommand {\gevBoldMath}  {\ensuremath{\mathrm{\pmb{GeV}}}}
\newcommand {\mmom}         {\ensuremath{\mathrm{MeV\kern-0.15em /\kern-0.12em c}}\xspace}
\newcommand {\gmom}         {\ensuremath{\mathrm{GeV\kern-0.15em /\kern-0.12em c}}\xspace}
\newcommand {\mmass}        {\ensuremath{\mathrm{MeV\kern-0.15em /\kern-0.12em c^2}}\xspace}
\newcommand {\gmass}        {\ensuremath{\mathrm{GeV\kern-0.15em /\kern-0.12em c^2}}\xspace}
\newcommand {\nb}           {\ensuremath{\mathrm{nb}}\xspace}
\newcommand {\musec}        {\ensuremath{\upmu \mathrm{s}}\xspace}
\newcommand {\nsec}         {\ensuremath{\mathrm{ns}}\xspace}
\newcommand {\psec}         {\ensuremath{\mathrm{ps}}\xspace}
\newcommand {\fmC}          {\ensuremath{\mathrm{fm}/c}\xspace}
\newcommand {\fm}           {\ensuremath{\mathrm{fm}}\xspace}
\newcommand {\mim}          {\ensuremath{\upmu\mathrm{m}}\xspace}
\newcommand {\cmq}          {\ensuremath{\mathrm{cm}^{2}}\xspace}
\newcommand {\mmq}          {\ensuremath{\mathrm{mm}^{2}}\xspace}
\newcommand {\dens}         {\ensuremath{\mathrm{g}/\mathrm{cm}^{3}}\xspace}
\newcommand {\lum}          {\ensuremath{\mathrm{cm}^{-2} \mathrm{s}^{-1}}\xspace}
\newcommand {\dg}           {\ensuremath{\kern+0.1em ^\circ}\xspace}
\newcommand{\mpp}           {\ensuremath{\mathrm{pp}}\xspace}
\newcommand{\rts}           {\ensuremath{\sqrt{s}}\xspace}
\newcommand{\gevc}          {\ensuremath{\mathrm{GeV}/c}\xspace}
\newcommand{\GeVc}          {\gevc}
\newcommand{\mevc}          {\ensuremath{\mathrm{MeV}/c}\xspace}
\newcommand{\mevcc}         {\ensuremath{\mathrm{MeV}/c^{2}}\xspace}
\newcommand{\gevcc}         {\ensuremath{\mathrm{GeV}/c^{2}}\xspace}
\newcommand{\pt}            {\ensuremath{p_\mathrm{T}}\xspace}
\newcommand{\kt}            {\ensuremath{k_\mathrm{T}}\xspace}
\newcommand{\mt}            {\ensuremath{m_\mathrm{T}}\xspace}
\newcommand {\lumi}         {\ensuremath{\mathcal{L}_\mathrm{int}}\xspace}
\newcommand{\nbinv}         {\ensuremath\mathrm{nb^{-1}}\xspace}
\newcommand {\ubinv}        {\ensuremath{\mu\rm b^{-1}}\xspace}
\newcommand{\mb}            {\ensuremath{\mathrm{mb}}\xspace}

\newcommand{\lt}{\textless}
\newcommand{\ctau}{\ensuremath{c\tau\xspace}}

%
%

\newcommand{\ePlusMinus}    {\ensuremath{\mathrm {e^{\pm}}}\xspace}
\newcommand{\muPlusMinus}   {\ensuremath{\upmu^{\pm}}\xspace}

\newcommand{\pion}          {\ensuremath{\uppi}\xspace}
\newcommand{\piZero}        {\ensuremath{\uppi^0}\xspace}
\newcommand{\piMinus}       {\ensuremath{\uppi^-}\xspace}
\newcommand{\piPlus}        {\ensuremath{\uppi^+}\xspace}
\newcommand{\piPlusMinus}   {\ensuremath{\uppi^\pm}\xspace}
\newcommand{\piMinusPlus}   {\ensuremath{\uppi^\mp}\xspace}

\newcommand{\proton}        {\ensuremath{\mathrm{p}}\xspace}
\newcommand{\pbar}          {\ensuremath{\mathrm{\overline{p}}}\xspace}
\newcommand{\pOuPbar}       {\ensuremath{\mathrm {p^\pm}}\xspace}
\newcommand{\Bminus}        {\ensuremath{\mathrm{B^-}}\xspace}
\newcommand{\Bplus}         {\ensuremath{\mathrm{B^+}}\xspace}
\newcommand{\Bpm}           {\ensuremath{\mathrm{B^\pm}}\xspace}
\newcommand{\Bzero}         {\ensuremath{\mathrm {B^0}}\xspace}
\newcommand{\Bzerobar}      {\ensuremath{\mathrm {\overline{B}^0}}\xspace}
\newcommand{\Bs}            {\ensuremath{\mathrm {B_s^0}}\xspace}
\newcommand{\Bsbar}         {\ensuremath{\mathrm {\overline{B}_s^0}}\xspace}

\newcommand{\rmLambdaZ}     {\ensuremath{\Lambda^0}\xspace}
\newcommand{\rmAlambdaZ}    {\ensuremath{\overline{\Lambda}^0}\xspace}
\newcommand{\rmLambda}      {\ensuremath{\Lambda}\xspace}
\newcommand{\rmAlambda}     {\ensuremath{\overline{\Lambda}}\xspace}
\newcommand{\rmLambdas}     {\ensuremath{\Lambda+\overline{\Lambda}}\xspace}

\newcommand{\Vzero}         {\ensuremath{\mathrm{V}^0}\xspace}
\newcommand{\Vzerob}        {\ensuremath{\mathrm{\pmb{V}}^0}\xspace}
\newcommand{\Kzero}         {\ensuremath{\mathrm{K}^0}\xspace}
\newcommand{\Kzs}           {\ensuremath{\mathrm{K_S}^0}\xspace}
\newcommand{\Ks}            {\Kzs}
\newcommand{\phimes}        {\ensuremath{\upphi}\xspace}
\newcommand{\Kminus}        {\ensuremath{\mathrm{K}^-}\xspace}
\newcommand{\Kplus}         {\ensuremath{\mathrm{K}^+}\xspace}
\newcommand{\Kstar}         {\ensuremath{\mathrm{K}^{*+}}\xspace}
\newcommand{\Kplusmin}      {\ensuremath{\mathrm{K}^\pm}\xspace}
\newcommand{\Jpsi}          {\ensuremath{\mathrm{J}/\uppsi}\xspace}
\newcommand{\DtoKpi}        {\ensuremath{\mathrm{D}^0 \to \mathrm{K}^-\uppi^+}\xspace}
\newcommand{\DbartoKpi}        {\ensuremath{\Dzerobar \to \mathrm{K}^+\uppi^-}\xspace}
\newcommand{\DtoKpipi}      {\ensuremath{\mathrm{D}^+\to \mathrm{K}^-\uppi^+\uppi^+}\xspace}
\newcommand{\DstartoDpi}    {\ensuremath{\Dstar \to \Dzero \uppi^+ \to \mathrm{K}^-\uppi^+\uppi^+}\xspace}
\newcommand{\Dstophip}      {\ensuremath{\mathrm{D_{s}^+ \to \upphi \uppi^+ \to K^+K^-\uppi^+}}\xspace}
\newcommand{\Dsminustophip}      {\ensuremath{\mathrm{D_{s}^- \to \upphi \uppi^- \to K^+K^-\uppi^-}}\xspace}
\newcommand{\Dzero}         {\ensuremath{\mathrm{D^0}}\xspace}
\newcommand{\Dzerobar}      {\ensuremath{\overline{\mathrm{D}}^0}\xspace}
\newcommand{\Dstar}         {\ensuremath{\mathrm{D^{*+}}}\xspace}
\newcommand{\Dstarm}        {\ensuremath{\mathrm{D^{*-}}}\xspace}
\newcommand{\Dplus}         {\ensuremath{\mathrm{D^+}}\xspace}
\newcommand{\Dminus}        {\ensuremath{\mathrm{D^-}}\xspace}
\newcommand{\Dpm}           {\ensuremath{\mathrm{D^\pm}}\xspace}
\newcommand{\Hcpm}          {\ensuremath{\mathrm{H_c^\pm}}\xspace}
\newcommand{\Ds}            {\ensuremath{\mathrm{D_{s}^+}}\xspace}
\newcommand{\Dsminus}       {\ensuremath{\mathrm{D_{s}^-}}\xspace}
\newcommand{\Dspm}          {\ensuremath{\mathrm{D_s^\pm}}\xspace}
\newcommand{\Lc}            {\ensuremath{\mathrm{\Lambda_{c}^+}}\xspace}
\newcommand{\Lcminus}       {\ensuremath\mathrm{{\overline{\Lambda}{}_c^-}}\xspace}
\newcommand{\Lcplus}        {\ensuremath\mathrm{{\Lambda_c^+}}\xspace}
\newcommand{\Lcpm}          {\ensuremath{\mathrm{\Lambda_{c}^\pm}}\xspace}
\newcommand{\Xic}           {\ensuremath{\Xi_\mathrm{c}}\xspace}
\newcommand{\lambdab}       {\Lb}
\newcommand{\lambdac}       {\Lc}
\newcommand{\xicz}          {\ensuremath{\Xi_\mathrm{c}^0}\xspace}
\newcommand{\xiczp}         {\ensuremath{\Xi_\mathrm{c}^{0,+}}\xspace}
\newcommand{\xicp}          {\ensuremath{\Xi_\mathrm{c}^+}\xspace}
\newcommand{\xib}           {\ensuremath{\Xi_\mathrm{b}^0}\xspace}
\newcommand{\Lbzero}        {\Lb}
\newcommand{\LctopKpi}      {\ensuremath{\mathrm{\Lambda_{c}^{+}\to p K^-\uppi^+}}\xspace}
\newcommand{\LbtoLc}        {\ensuremath{\mathrm{\Lb \to \Lc + \mathrm{X}}}\xspace}
\newcommand{\LctopKzeros}   {\ensuremath{\mathrm{\Lc \to p \Kzs}}\xspace}
\newcommand{\KzStopippim}   {\ensuremath{\mathrm{\Kzs \to \uppi^+ \uppi^-}}\xspace}
\newcommand{\Lambdatoppim}  {\ensuremath{\mathrm{\Lambda \to p \uppi^{-}}}\xspace}

\newcommand{\decleng}       {\ensuremath{\mathrm{L_{xyz}}}\xspace}
\newcommand{\cosP}          {\ensuremath\mathrm{cos_{\Theta_{pointing}}}\xspace}

\newcommand{\ptLc}          {\ensuremath{p_\mathrm{T, \Lambda_c}}\xspace}
\newcommand{\ptpion}        {\ensuremath{p_\mathrm{T, \uppi}}\xspace}
\newcommand{\ptK}           {\ensuremath{p_\mathrm{T, K}}\xspace}
\newcommand{\ptproton}      {\ensuremath{p_\mathrm{T, \proton}}\xspace}

\newcommand{\da}            {\partial}
\newcommand{\de}            {\mathrm{d}}
\newcommand{\slfrac}[2]     {\left.#1\right/#2}
\newcommand{\av}[1]         {\left\langle #1 \right\rangle}
\newcommand{\Gevc}          {\gevc}
\newcommand{\GeV}           {\gev}
\newcommand{\MeV}           {\mev}
\newcommand{\MeVcc}         {\mevcc}
\newcommand{\TeV}           {\tev}
\newcommand{\dm}            {\ensuremath{\Delta M}\xspace}

\newcommand{\mub}           {\ensuremath{\mathrm{\upmu b}}\xspace}
\newcommand{\pb}            {\ensuremath{\mathrm{pb}}\xspace}
\newcommand{\mum}           {\ensuremath{\mathrm{\upmu m}}\xspace}
\newcommand{\DzerotoKpi}    {\DtoKpi}
\newcommand{\DplustoKpipi}  {\DtoKpipi}
\newcommand{\DminustoKpipi}  {\ensuremath{\mathrm{D}^-\to \mathrm{K}^+\uppi^-\uppi^-}\xspace}
\newcommand{\Dstophipi}     {\ensuremath{\mathrm{D_s^{+}\to \upphi\uppi^+}}\xspace}
\newcommand{\DstophipitoKKpi} {\ensuremath{\mathrm{D_s^{+}\to [\upphi\to\mathrm{K^+K^-}] \uppi^+}}\xspace}
\newcommand{\phitoKK}       {\ensuremath{\mathrm{\upphi\to  K^+K^-}}\xspace}
\newcommand{\KKpi}          {\ensuremath{\mathrm{K^-\uppi^+\uppi^+}}\xspace}
\newcommand{\RAA}           {\ensuremath{R_\mathrm{AA}}\xspace}
\newcommand{\Ntrkl}         {\ensuremath{N_\mathrm{trkl}}\xspace}
\newcommand{\averNtrkl}     {\ensuremath{\langle N_\mathrm{trkl} \rangle}\xspace}
\newcommand{\Nch}           {\ensuremath{N_\mathrm{ch}}\xspace}
\newcommand{\averNch}       {\ensuremath{\langle N_\mathrm{ch} \rangle}\xspace}
\newcommand{\zVtx}          {\ensuremath{z_\mathrm{vtx}}\xspace}
\newcommand*\code[1]        {\texttt{#1}}
\newcommand{\fprompt}       {\ensuremath{f_\mathrm{prompt}}\xspace}
\newcommand{\fnonprompt}    {\ensuremath{f_\mathrm{non\text{-}prompt}}\xspace}
\newcommand{\AccEff}        {\ensuremath{\mathrm{Acc}\times\epsilon}\xspace}
\newcommand{\effP}[1]       {\ensuremath{\epsilon_#1^\mathrm{p}}\xspace}
\newcommand{\effNP}[1]      {\ensuremath{\epsilon_#1^\mathrm{np}}\xspace}
\newcommand{\rawY}[1]       {\ensuremath{Y_#1}\xspace}
\newcommand{\fP}            {\ensuremath{f_\mathrm{p}}\xspace}
\newcommand{\fNP}           {\ensuremath{f_\mathrm{np}}\xspace}

\newcommand{\HbtoDstar}     {\ensuremath{\mathrm{H_b\to\Dstar+X}}\xspace}
\newcommand{\fbtoHb}        {\ensuremath{f(\mathrm{b\to H_b})}\xspace}
\newcommand{\fbtoBzero}        {\ensuremath{f(\mathrm{b\to\Bzero})}\xspace}
\newcommand{\bbbar}         {\ensuremath{\mathrm{b\overline{b}}}\xspace}

\newcommand{\rzz}           {\ensuremath{\uprho_{00}}\xspace}
\newcommand{\rzzp}          {\ensuremath{\uprho_{00}^\mathrm{prompt}}\xspace}
\newcommand{\rzznp}         {\ensuremath{\uprho_{00}^\mathrm{non\text{-}prompt}}\xspace}
\newcommand{\rzznpevtg}     {\ensuremath{\uprho_{00}^\mathrm{non\text{-}prompt,\evtgen}}\xspace}
\newcommand{\cost}          {\ensuremath{\cos{\vartheta^*}}\xspace}
\newcommand{\costsq}        {\ensuremath{\cos^2{\vartheta^*}}\xspace}
\newcommand{\costHel}       {\ensuremath{\cos{\vartheta^*}_\mathrm{helicity}}\xspace}
\newcommand{\costProd}      {\ensuremath{\cos{\vartheta^*}_\mathrm{production}}\xspace}
\newcommand{\ZZero}         {\ensuremath{\mathrm{Z}^0}\xspace}
\newcommand{\Lb}            {\ensuremath{\mathrm{{\Lambda_b^0}}}\xspace}
\newcommand{\pythia}        {\textsc{Pythia~8}\xspace}
\newcommand{\herwig}        {\textsc{Herwig~7}\xspace}
\newcommand{\geant}         {\textsc{Geant~4}\xspace}
\newcommand{\evtgen}        {\textsc{EvtGen}\xspace}
\newcommand{\crblc}         {\textsc{Crblc~Mode 2}\xspace}

\newcommand{\BzerotoDmpi}          {\ensuremath{\Bzero\to\Dminus\uppi^+}\xspace}
\newcommand{\BzerotoDmpitoPikpi}   {\ensuremath{\Bzero\to \Dminus\left[\to\mathrm{K^+\uppi^-\uppi^-}\right]\uppi^+}\xspace}
\newcommand{\BplustoDzpi}        {\ensuremath{\Bplus\to\Dzerobar\uppi^+}\xspace}  
\newcommand{\BplustoDzpitokpi}   {\ensuremath{\Bplus\to\Dzerobar\left[\to\mathrm{K^+\uppi^-}\right]\uppi^+}\xspace}
\newcommand{\BstoDspi}   {\ensuremath{\Bs\to\Dsminus\uppi^+}\xspace}
\newcommand{\BstoDspitokkpi}   {\ensuremath{\Bs\to\Dsminus\left[\to\upphi\uppi^-\to\mathrm{K^+K^-\uppi^-}\right]\uppi^+}\xspace}
\newcommand{\LbtoLcpi}          {\ensuremath{\Lb\to\Lc\uppi^-}\xspace}

\newcommand{\muR}           {\ensuremath{\mu_\mathrm{R}}\xspace}
\newcommand{\muF}           {\ensuremath{\mu_\mathrm{F}}\xspace}
\newcommand{\muRF}           {\ensuremath{\mu_\mathrm{R,F}}\xspace}
\newcommand{\nnlonnll}{NNLO+NNLL\xspace}

\begin{titlepage}
\PHyear{2026}       
\PHnumber{077}      
\PHdate{13 March}  

\title{Measurement of the $\mathbf{B^0}$-meson production cross section in proton--proton collisions at $\mathbf{\sqrt{\textit{s}}=13.6}$ TeV}
\ShortTitle{\Bzero cross section in pp collisions at $\sqrts=13.6~\TeV$}   
%
\Collaboration{ALICE Collaboration\thanks{See Appendix~\ref{app:collab} for the list of collaboration members}}
\ShortAuthor{ALICE Collaboration} 

\begin{abstract}
This article reports the measurement of the transverse-momentum ($p_\mathrm{T}$) differential production cross section of $\mathrm{B^0}$ mesons in proton--proton collisions at a centre-of-mass energy of $\sqrt{s}=13.6~\mathrm{TeV}$ with the ALICE detector at the CERN LHC. For the first time, the $\mathrm{B^0}$ production cross section is measured at midrapidity ($|y|<0.5$) down to $p_\mathrm{T}=1~\mathrm{GeV}/c$ at LHC energies. The $\mathrm{B^0}$ mesons and their charge conjugates were reconstructed via the $\mathrm{B^0}\to\mathrm{D^{-}}\uppi^+$ decay channel, followed by the $\mathrm{D^-}\to\mathrm{K^+\uppi^-\uppi^-}$ decay. The measured $p_\mathrm{T}$-differential production cross section is described within uncertainties by state-of-the-art models based on perturbative quantum-chromodynamics calculations. Its rapidity dependence is also studied by computing the $p_\mathrm{T}$-differential ratios between the ALICE measurement and the one of $\mathrm{B^+}$ mesons performed by the LHCb Collaboration at forward rapidity. The $\mathrm{B^0}$ production cross section per unit of rapidity at midrapidity is  $\mathrm{d}\sigma(\mathrm{B^0})/\mathrm{d} y|_{|y|<0.5} = 24.2 \pm 1.4~(\text{stat.}) \pm 2.6~(\text{syst.})_{-0.3}^{+0.2}~(\text{extrap.})~\mathrm{\upmu b}$.
\end{abstract}
\end{titlepage}

\setcounter{page}{2}

%
\section{Introduction}
\label{sec:intro}
Measurements of the production cross section of beauty hadrons in proton--proton (pp) collisions provide a sensitive test of perturbative quantum-chromodynamics (pQCD) calculations.
Theoretical approaches based on factorisation theorems implement the calculation of the transverse momentum ($\pt$) and rapidity ($y$) differential hadron production cross sections as the convolution of three components: (i) the parton distribution functions (PDFs) of the colliding protons; (ii) the partonic scattering cross section, expressed as a perturbative series expansion in the strong coupling constant; (iii) the fragmentation function (FF), which describes the non-perturbative transition of the beauty quark into a given hadron.
The FF is typically parameterised using measurements from \ee or ep collisions, under the assumption that the hadronisation process is universal, i.e., independent of the collision system.
Factorisation can be implemented in terms of the transferred momentum squared $Q^2$ (collinear factorisation)~\cite{Collins:1989gx} or the parton transverse momentum $k_{\rm T}$~\cite{Catani:1990eg}.
Within the collinear-factorisation approach, calculations with the General-Mass Variable-Flavour-Number Scheme (GM-VFNS)~\cite{Kniehl:2007erq,Benzke:2019usl,Helenius:2023wkn} and with the Fixed Order plus Next-to-Leading Logarithms (FONLL)~\cite{Cacciari:2012ny} frameworks have been the standard baseline for over 20 years. They achieve next-to-leading order (NLO) accuracy with all-order resummation of next-to-leading logarithms. This is accomplished by matching a massless-quark calculation with resummed logarithmic terms, valid at high $\pt$, to a fixed-order calculation with massive quarks, accurate in the low-$\pt$ region. 
In parallel, calculations in the $k_{\rm T}$-factorisation framework were developed~\cite{Shabelski:2004qy,Shabelski:2017kmy,Maciula:2019izq}. The most recent predictions with this approach employ the variable-flavour-number scheme approach and include all relevant NLO contributions~\cite{Guiot:2021vnp,Barattini:2025wbo}.
Beyond NLO, fully differential calculations of beauty-quark production at next-to-next-to-leading-order (NNLO) in the collinear factorisation approach have recently been published~\cite{Catani:2020kkl,Mazzitelli:2023znt}. The NNLO predictions for beauty-quark production in pp collisions at Large Hadron Collider (LHC) energies overlap well with the NLO uncertainty bands, with a significant reduction in the uncertainties related to the perturbative expansion, estimated through variations of scale parameters, thus suggesting convergence of the perturbative series. More recently, NNLO calculations that include the resummation of collinear logarithms at next-to-next-to-leading log (NNLL) accuracy and an appropriate non-perturbative fragmentation function to model the quark-to-hadron transition have become available~\cite{Czakon:2024tjr}, and are characterised by smaller uncertainties as compared to predictions at lower perturbative accuracy. The resummation has a significant impact at high \pt, while the inclusion of the fragmentation function enables direct comparisons with measurements of beauty-hadron production cross sections.

The relative production rates of the different beauty-hadron species, commonly denoted as fragmentation fractions, provide insights into the beauty-quark hadronisation process. Measurements of meson-to-meson production ratios for both charm and beauty mesons in pp and p–nucleus collisions at the LHC have shown consistency with earlier measurements in \ee and ep collisions~\cite{Altmann:2024kwx}. However, heavy-flavour baryon measurements in hadronic collisions at LHC energies have revealed a significant increase in charm- and beauty-baryon production relative to mesons, compared to the \ee baseline~\cite{ALICE:2022wpn,ALICE:2023sgl,LHCb:2019fns,LHCb:2023wbo,ALICE:2023wbx}. This challenges the assumption that heavy-quark hadronisation is a universal process across different collision systems. Various models have been proposed to describe the enhanced charm- and beauty-baryon yields, such as an extension of colour reconnection in string fragmentation models beyond the leading-colour approximation~\cite{Christiansen:2015yqa}, the inclusion of a hadronisation mechanism via quark coalescence~\cite{Minissale:2020bif,Beraudo:2023nlq,Zhao:2024ecc}, or the addition of a set of yet-unobserved higher-mass heavy-flavour baryons predicted by the relativistic-quark model (RQM~\cite{Ebert:2011kk}) in a statistical hadronisation approach~\cite{He:2019tik,He:2022tod} (see~\cite{Altmann:2024kwx} for a recent review).

Various measurements of beauty production have been performed in pp collisions at the LHC for centre-of-mass energies (\sqrts) ranging from 2.76 to 13 TeV. Beauty production has been studied through measurements of non-prompt J/$\uppsi$ and non-prompt D mesons~\cite{ALICE:2021edd,ALICE:2021mgk,ALICE:2024xln,ALICE:2023wbx,ATLAS:2015zdw,LHCb:2015foc}, beauty-decay leptons~\cite{ALICE:2012acz,CMS:2011xhf}, dielectrons~\cite{ALICE:2020mfy,ALICE:2018gev}, b-tagged jets~\cite{ATLAS:2011ac,CMS:2012pgw}, and partially reconstructed semileptonic decays ($\mathrm{H_b \rightarrow H_c \upmu X}$, where $\mathrm{H_b}$ and $\mathrm{H_c}$ indicate beauty and charm hadrons, respectively)~\cite{ATLAS:2012sfc,LHCb:2016qpe,LHCb:2019fns}. Fully reconstructed decays of beauty hadrons have been measured by the ATLAS~\cite{ATLAS:2013cia} and CMS~\cite{CMS:2011pdu,CMS:2016plw,CMS:2024vip} Collaborations at midrapidity in the high-$\pt$ region ($\pt>10~\GeV/c$ at $\sqrts=5.02$ and 13~\TeV, $\pt>5~\GeV/c$ at $\sqrts=7~\TeV$) and at forward rapidity by the LHCb Collaboration~\cite{LHCb:2017vec} down to $\pt=0$.
At lower collision energies, measurements have been performed in pp collisions at $\sqrts = 200~\GeV$ at RHIC~\cite{STAR:2010ibw,PHENIX:2019pxh} and in $\mathrm{p\overline{p}}$ collisions at $\sqrts = 0.63~\TeV$ at the S$\mathrm{p\overline{p}}$S~\cite{UA1:1988iqx,UA1:1990vvp} and at $\sqrts = 1.8$ and $1.96~\TeV$ at the Tevatron~\cite{CDF:2004jtw,CDF:2006ipg,CDF:2009bqp,D0:1994vpd}.
The measured cross sections are generally described within uncertainties by pQCD-based calculations.

In this article, we present the measurement of the \pt-differential production cross section of fully-reconstructed $\Bzero$ mesons in pp collisions at $\sqrts=13.6~\TeV$. The \Bzero production is measured at midrapidity ($|y|<0.5$) in the transverse momentum interval $1 < \pt < 23.5~\GeV/c$. This is the first measurement of B-meson production at low $\pt$ in the midrapidity region at LHC energies, significantly extending the low-$\pt$ reach of previous B-meson production measurements at midrapidity at $\sqrts=$ $13~\TeV$ by the CMS Collaboration~\cite{CMS:2016plw}. The results reported here are also complementary to those obtained for \Bplus mesons at forward rapidity ($2<y<4.5$) at $\sqrts=$ $13~\TeV$ by the LHCb Collaboration~\cite{LHCb:2017vec}, which cover a wide transverse momentum interval ($0<\pt<40~\GeV/c$). These new measurements provide a test of pQCD calculations of beauty production in a kinematic region where direct measurements of beauty hadrons were missing at LHC energies.

\section{Experimental apparatus and data samples}
\label{sec:exp}

\sloppy The ALICE experimental setup consists of two primary components: the central barrel and the muon spectrometer, designed to cover the pseudorapidity region $|\eta| < 0.9$ and  \mbox{$-4 < \eta < -2.5$}, respectively~\cite{ALICE:2014sbx}. The central barrel detectors are housed within the L3 magnet, a large solenoid generating a uniform magnetic field of 0.5\,T aligned with the beam axis.

The ALICE detector has successfully undergone a major upgrade during the second LHC long shutdown (2019--2021)~\cite{ALICE:2023udb}. 
This upgrade includes the installation of a new innermost ALICE detector, the Inner Tracking System (ITS), the replacement of the multi-wire proportional chambers used for the readout of the Time Projection Chamber (TPC) detector with ones based on gas electron multipliers (GEM)~\cite{ALICETPC:2020ann}, and the implementation of new readout electronics for the Time-Of-Flight (TOF) detector~\cite{ALICETOF:2025tof}. The new ITS consists of seven layers of silicon detectors based on monolithic active pixel sensors~\cite{ALICE:2013nwm}, allowing for a precise determination of the track parameters in the vicinity of the interaction point, and consequently of the positions of interaction (primary) and decay (secondary) vertices. The TPC provides up to 152 space points to reconstruct the charged-particle trajectory, as well as particle identification via the measurement of the specific ionisation energy loss \dEdx. The particle identification capabilities of the TPC are extended by the TOF detector, which is used to measure the flight time of charged particles from the interaction point. A new Fast Interaction Trigger (FIT) detector suite~\cite{Trzaska:2017reu}, consisting of the FV0, FT0, and FDD sub-detectors, was installed in the forward and backward regions, and serves as a luminosity counter as well as a centrality estimator in heavy-ion collisions. 

With the detector upgrades implemented before the start of the LHC Run 3, in particular the GEM–based TPC readout chambers, it is possible to operate the detector in a continuous data-taking scheme~\cite{ALICE:2023udb,Buncic:2015ari}. 
The new readout system allows operation at an interaction rate of 660 kHz and higher in pp collisions, significantly larger than the one reached during the LHC Run 2, while maintaining excellent tracking and particle identification performance. Instead of relying on traditional hardware triggers to select individual events, data are collected in time frames (TFs) of $\approx3$ milliseconds, corresponding to 32 LHC orbits, which contain all detector signals accumulated during this interval~\cite{ALICE:2023udb, Buncic:2015ari, Lettrich:2752837}. Trigger and event selection are performed in software after full calibration and reconstruction within the $O^2$ (Online–Offline) computing framework~\cite{Buncic:2015ari}. 
To separate data from different collisions occurring within a time frame and to mitigate the effects of out-of-time pileup, where signals from preceding or following collisions may overlap within a common time interval of the TFs due to the detector response time, precise timing information from detectors such as the FT0 and the high-resolution tracking provided by the upgraded ITS are exploited. However, due to finite time resolution, a fraction of reconstructed tracks may still be compatible with multiple primary vertices. To manage this ambiguity, the reconstruction
process evaluates the compatibility of tracks with vertices based on timing and spatial proximity. Moreover, the decay reconstruction and the selection of specific decay-vertex topologies strongly suppress the fraction of tracks associated to the wrong vertices, as discussed in the following.

A dedicated event-skimming strategy is employed to reduce data volume. All minimum-bias events are first buffered and reconstructed, and then only those events that meet predefined physics criteria, such as the presence of a beauty-hadron candidate required for this analysis, are retained. To accommodate the drift time in the TPC from the central electrode to the readout chambers, all events within a time window of $-25~\musec$ and $+125~\musec$ around each selected event are also preserved. In this way, efficient data selection is enabled while preserving all relevant physics information, with about 4.5\% of the original data volume retained after event-skimming~\cite{ALICE-PUBLIC-2020-005}.

The data sample of pp collisions at $\sqrts=13.6~\TeV$ used for this analysis was collected in the years 2023 and 2024.  The integrated luminosity \lumi was determined from visible cross sections measured in van der Meer (vdM) scans, similarly to what was reported in Ref.~\cite{ALICE:2021leo}. In the vdM scans, the two beams are moved across each other in the transverse (horizontal and vertical) directions. The scans in the two directions are performed separately, the beams being head-on in the non-scanned direction. Measurement of the rate for a given visible process as a function of the beam separation allows the determination of the luminosity for head-on collisions, and of the visible cross section. The latter was measured for an FT0-based minimum-bias trigger, which was used as the main luminosity signal, and for a similar, FDD-based, trigger. A conservative uncertainty of 3\% and 4\% was assigned for the integrated luminosity of the 2023 and 2024 datasets, respectively, and used for the results reported in the following. These uncertainties are dominated by interaction-rate-dependent differences between the FT0- and FDD-based luminosities. The total integrated luminosity was determined to be $\lumi = 48.5 \pm 1.9~\pb^{-1}$, after considering a correction of 1.2\% for the in-bunch pileup, resulting from an average value of visible interactions per bunch crossing of about $\mu_\mathrm{vis}=2.36\%$.

The recorded collisions were required to have a time signal from the FIT detector that corresponds to a bunch crossing time provided by the LHC.
Only collisions with a reconstructed primary vertex within 10 cm of the nominal position along the beam axis were selected.
The offline trigger selection (OTS) strategy for the analysis described in this article was developed to select events containing decays of beauty hadrons into a D meson and a pion. Ground-state beauty hadrons decay weakly with a mean proper decay length ($c\tau$) between 441 and 491 \mum, while the charm hadrons produced in their decay have $c\tau$ values ranging from 66 to 321~\mum. As a result, charm hadrons from beauty decays, referred to as non-prompt in the following, are more displaced from the primary vertex than prompt charm hadrons, originating from charm-quark hadronisation or from decays of promptly-produced excited charm-hadron or charmonium states.

The decay channel considered was \BzerotoDmpi, with a branching ratio (BR) of $(2.51 \pm 0.08) \times 10^{-3}$, followed by the \Dminus-meson decay into the \DminustoKpipi channel with $\text{BR}=(9.38 \pm 0.16) \times 10^{-2}$~\cite{ParticleDataGroup:2024cfk}, together with its charge conjugate. 
The invariant mass of the combination of the reconstructed \Dpm and charged pion was required to be compatible within $400~\MeVcc$ with the PDG world-average mass of the \Bzero meson~\cite{ParticleDataGroup:2024cfk}.
\Dplus-meson candidates were built by combining triplets of tracks with the proper charge signs, each with $|\eta| < 0.8$, $\pt > 0.3~\gev/c$, at least 70 (out of 152) associated space points in the TPC, and a minimum of four (out of seven) hits in the ITS, with at least one in either of the three innermost layers to ensure a good pointing resolution.
They were further selected adopting a machine-learning approach based on Boosted Decision Trees (BDT) as provided by the XGBoost library~\cite{Chen:2016XST,hipe4ml}. Signal samples of prompt and non-prompt \Dminus mesons, as well as background candidates for the BDT training, were obtained from Monte Carlo (MC)  simulations. Before the training, loose kinematic and topological selections were applied together with the particle identification of decay-product tracks.
Pions and kaons were selected by requiring both the measured TPC \dEdx and time of flight to be compatible with the signal expected for the relevant particle hypothesis within three standard deviations ($3\,\sigma$). Tracks without TOF hits were identified using only the TPC information.
The features provided to the BDTs as input to distinguish among prompt charm hadrons, non-prompt ones, and combinatorial background were based on the track \pt and their displacement from the primary vertex on the transverse plane ($d_0^{xy}$) as well as along the beam axis ($d_0^z$), exploiting the different mean proper decay lengths of charm and beauty hadrons. 
The BDT outputs are related to the candidate probability to be a prompt or non-prompt \Dminus meson, or combinatorial background. The candidates were required to have a high probability of being a non-prompt \Dminus  and a low probability of being a combinatorial-background candidate.
The track associated with the charged pion, to be paired with the \Dminus -meson candidate, was required to have $\pt>1~\GeV/c$ and $|d_0^{xy}|>50~\mum$. 

To assess possible inefficiencies introduced by the further reconstruction of raw data selected by the OTS, dedicated downscaled triggers requiring only the presence of a non-prompt charm hadron fulfilling the conditions described above were used.
The efficiency was estimated as the ratio between the raw signal yield of charm hadrons in the dedicated trigger sample and the minimum-bias one. In both samples, the same selection criteria used for the analysis were applied. The yields were extracted by fitting the invariant-mass distributions and normalised by the respective luminosities. The minimum-bias sample corresponds to 0.5\% of the total integrated luminosity inspected. The invariant-mass fits were performed using a double-sided Crystalball function to describe the signal, and an exponential function for the combinatorial background. A systematic uncertainty ranging from 2\% to 9\%, from low to high D-meson \pt, was assigned to the signal extraction. It was estimated by varying the fit range and the functional form of the combinatorial background, as well as by integrating the invariant-mass distribution after background subtraction instead of using the signal fit function.

\begin{figure}[!tb]
  \begin{center}
     \vspace{3ex}
     \includegraphics[width=0.58\linewidth]{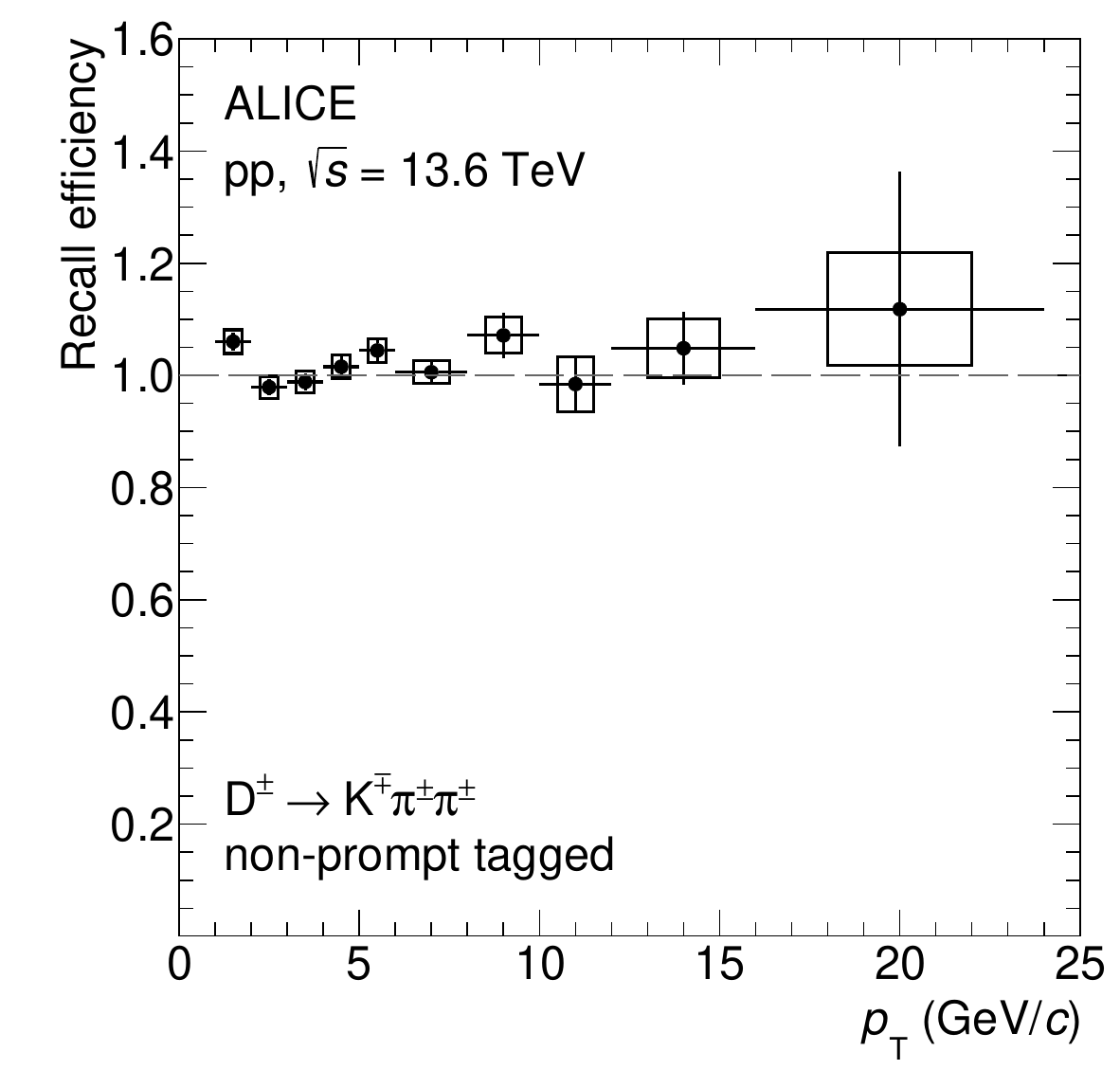}
     \caption[Toc caption]{\label{fig:triggerEff}Recall efficiency for non-prompt \Dpm mesons estimated as the ratio between the yield from the offline-trigger-selected dataset and the yield extracted by applying the same selection criteria to a minimum-bias data sample, as described in the text.}
  \end{center}
\end{figure}

The measured yield ratio, referred to as ``recall efficiency" is reported in Fig.~\ref{fig:triggerEff} for the sample of \Dpm mesons tagged as non-prompt, as described above, and shows that the offline selections are fully efficient within uncertainties. 
The recall efficiency demonstrates that the further reconstruction of raw data after the OTS does not introduce any additional inefficiency in the reconstruction and selection of non-prompt \Dpm mesons compared to an analysis on minimum-bias data. The recall efficiency was evaluated for the \Dpm mesons, since the size of the minimum-bias sample did not allow for a verification of \Bzero signals directly; however, the request of a \Dpm-pion pair is not expected to alter the conclusion.

The Monte Carlo samples of pp collisions used in this analysis for the efficiency correction and the training of machine-learning models were produced with the PYTHIA 8.3~\cite{Bierlich:2022pfr} generator with colour-reconnection beyond-leading-colour approximation (CRBLC) Mode 2~\cite{Christiansen:2015yqa}. The simulated events were required to contain a \bbbar pair, and the beauty hadrons were forced to decay in the decay channels of interest in the analysis. These also include channels that contribute to the correlated backgrounds in the invariant-mass distributions, as described in the next section. The generated particles were propagated through the detector using the GEANT4 transport package~\cite{GEANT4:2002zbu}. The luminous region distribution, as well as the detailed conditions of all ALICE detectors, including active channel configurations, gain, and noise levels, were taken into account in the simulations, considering also their time-dependent evolution throughout data taking.
The spatial resolution on the track position at the primary vertex was further tuned in the MC simulations to match that estimated on the data.

\section{Data analysis}
\label{sec:analysis}

The analysis of OTS events started with the reconstruction of the signal candidates using the same selection criteria as those described in Section~\ref{sec:exp}. To further reduce the combinatorial background and preserve the highest possible efficiency for the \Bzero-meson selection, a second binary classification was employed via BDTs trained using features related to the \Bzero decay-vertex topology. For the BDT training, signal samples of beauty mesons were obtained from MC simulations, while background samples were obtained from data using candidates with invariant mass in the intervals $4.9 < M(\mathrm{D^\mp\uppi^\pm}) < 5.0$~\gevcc and $5.56 < M(\mathrm{D^\mp\uppi^\pm}) < 5.66$~\gevcc, on both sides away from the \Bzero meson nominal mass~\cite{ParticleDataGroup:2024cfk}.
Independent BDT models were trained in the different \Bzero-meson \pt intervals. The BDT output score is related to the candidate's probability of being signal or background. Selections on the signal BDT score were optimised to obtain high efficiency while discarding enough background to maintain a good level of expected statistical significance. The latter was calculated using an estimated value for the signal from FONLL predictions, multiplied by the reconstruction and selection efficiencies for each considered BDT selection threshold, and an estimate of the background within the signal region obtained by interpolating a fit to the invariant-mass distribution in the sidebands of the \Bzero signal region. The applied selections also ensured the suppression of the signal candidates associated with the wrong primary vertex due to time ambiguities (see Section~\ref{sec:exp}) to a fraction smaller than 0.1\%.

\begin{figure}[!tb]
  \begin{center}
     \includegraphics[width=0.96\linewidth]{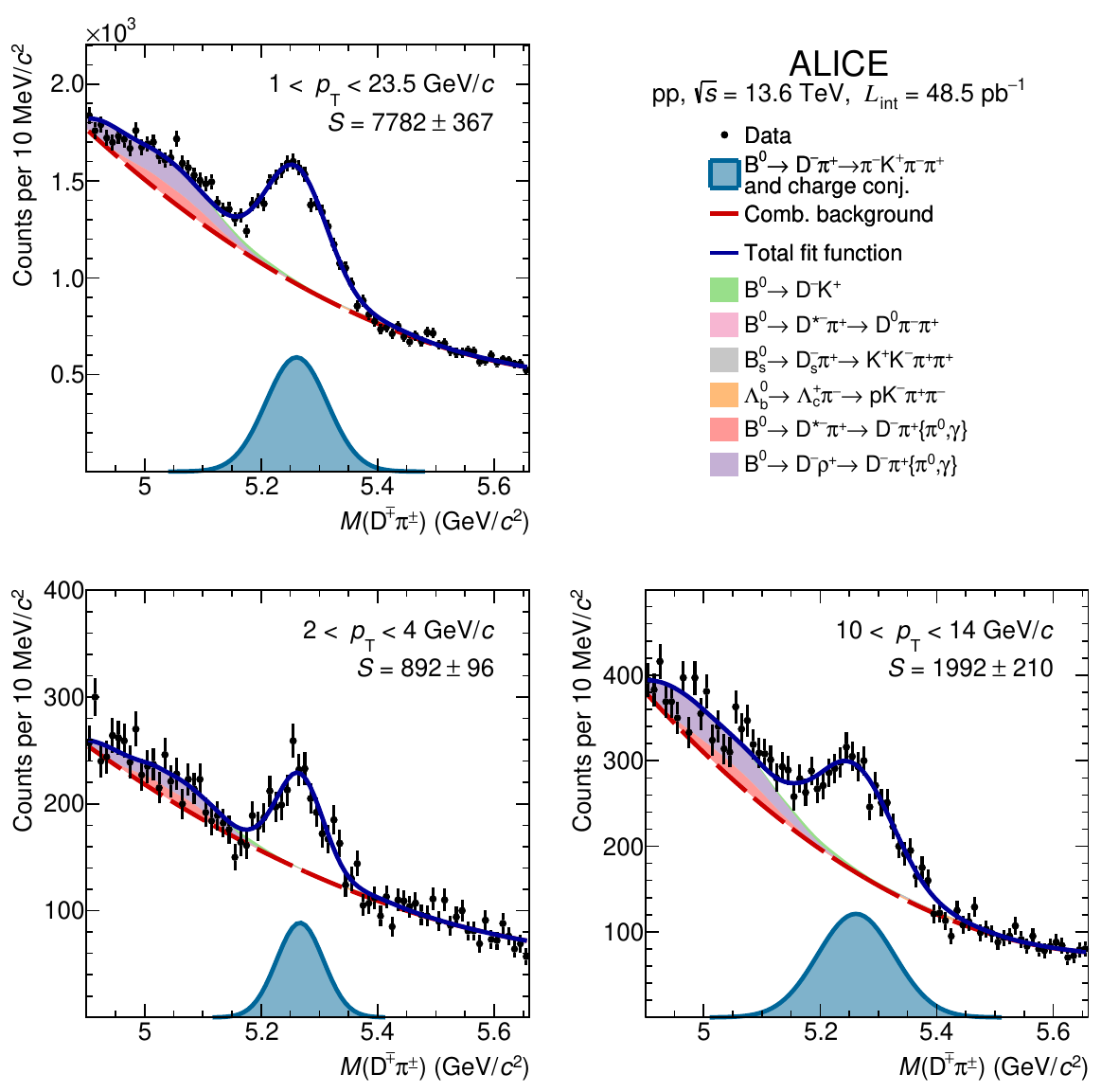}
     \caption[Toc caption]{\label{fig:raw_yields} Invariant-mass distributions of \Bzero-meson candidates in the $1 < \pt < 23.5 ~ \gevc$ (top), $2 < \pt < 4 ~ \gevc$ (bottom-left), and $10 < \pt < 14 ~ \gevc$ (bottom-right) intervals. The contributions of the different partially and mis-reconstructed decays to the total fit function (blue) are reported on top of the combinatorial background fit function (red dashed line) with filled colour-coded areas. The values of the signal counts ($S$) are reported in the text. }
  \end{center}
\end{figure}

The raw yields of \mbox{$\Bzero \to \Dminus \left[ \to \Kplus\uppi^-\uppi^-\right] \uppi^+$} mesons, including both particles and antiparticles, were extracted via unbinned maximum-likelihood fits to the invariant-mass distributions of the selected beauty-meson candidates~\cite{Eschle:2019jmu, grosa_2023_13902597}, performed in seven transverse-momentum intervals in the range $1 < \pt < 23.5 ~ \gevc$. The fitting function was composed of a second-order polynomial to describe the combinatorial background and a Gaussian term for the signal. Additional background sources arising from other \Bzero decays such as \mbox{$\Bzero \to \Dstarm \left[ \to \Dminus ~ \uppi^0/\gamma\right] \uppi^+$}, \mbox{$\Bzero \to\Dminus\rho^+\left[\to\uppi^+~ \uppi^0/\gamma\right]$}, or \mbox{$\Bzero \to \Dminus \mathrm{K}^+$}, or from decays of other beauty hadrons, such as \mbox{$\Lb \to \Lc\left[ \to \mathrm{p} \Kminus \uppi^+\right] \uppi^-$}, were considered. These candidates from partially and mis-reconstructed decays contribute to the fit function via the inclusion of templates obtained from the reconstructed candidates in MC simulations and modelled with a kernel density estimate~\cite{Parzen:Kde, Rosenblatt:Kde}. The normalisation of each contribution arising from a \Bzero decay was fixed relative to the signal using the branching ratios reported in Ref.~\cite{ParticleDataGroup:2024cfk}. The contributions arising from decays of \Bs and \Lb were normalised in a similar way, using PYTHIA 8.3~\cite{Bierlich:2022pfr} with CRBLC Mode 2~\cite{Christiansen:2015yqa} predictions for their expected relative production rates with respect to the \Bzero meson. While the \mbox{$\Bzero \to \Dstarm \left[ \to \Dminus ~ \uppi^0/\gamma\right] \uppi^+$} and \mbox{$\Bzero \to\Dminus\rho^+\left[\to\uppi^+~ \uppi^0/\gamma\right]$} correlated backgrounds contribute roughly $20\text{--}30\%$ relative to the signal yield, and the relative contributions of \mbox{$\Lb \to \Lc\left[ \to \mathrm{p} \Kminus \uppi^+\right] \uppi^-$} and \mbox{$\Bzero \to \Dminus \Kminus$} decays are about $2\text{--}4\%$, those of \mbox{$\Bzero \to \Dstarm[\to\Dzero\uppi^-]\uppi^+$} and \mbox{$\Bs \to \Dsminus[\to\Kplus\Kminus\uppi^+]\uppi^+$} are significantly smaller, on the order of few permil. The invariant-mass distributions, together with the result of the fits, are reported in \figref{fig:raw_yields} for the intervals $1 < \pt < 23.5 ~ \gevc$, $2 < \pt < 4 ~ \gevc$, and $10 < \pt < 14 ~ \gevc$.

The \pt-differential production cross section of \Bzero mesons at midrapidity was computed as:
\begin{equation}
\left. \cfrac{\mathrm{d}^2 \sigma(\Bzero)}{\dpt \dy} \right|_{|y|<0.5}
=
\cfrac{1}{2} \times
\cfrac{N^{\Bzero + \Bzerobar}_{\rm raw} (\pt)}{(\mathrm{Acc} \times \varepsilon) (\pt)\times c_{\Delta y}(\pt)} \times \cfrac{1}{\Delta \pt \times \Delta y  \times\mathrm{BR} \times \lumi} ~ ,
\end{equation}

where $N^{\Bzero + \Bzerobar}_{\rm raw}$ represents the raw yields extracted in each \pt interval, and the factor $1/2$ is included to account for the fact that the raw yields contain both particles and antiparticles, while the production cross section is given as an average of particles and antiparticles. The yields were divided by the width of the \pt interval ($\Delta \pt$), the width of the rapidity interval ($\Delta y=1$), the acceptance times efficiency $(\mathrm{Acc} \times \varepsilon)$, the correction factor for the rapidity coverage $c_{\Delta y}$ as defined in the next paragraph, the \mbox{$\mathrm{BR} = (2.35 \pm 0.08) \times 10^{-4}$}~\cite{ParticleDataGroup:2024cfk} of the full decay channel, and the integrated luminosity \lumi.

\begin{figure}
    \centering
    \includegraphics[width=0.5\linewidth]{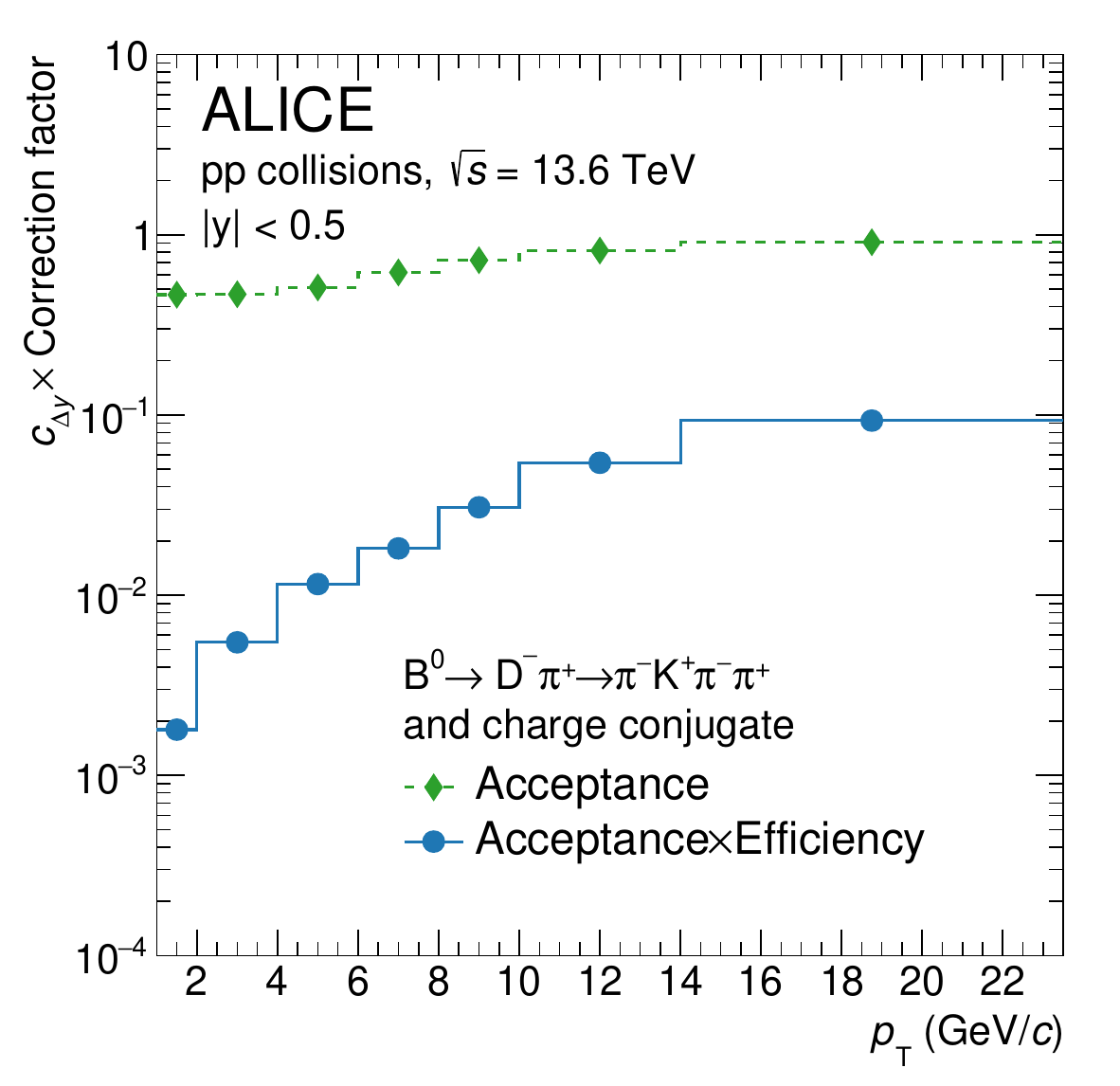}
    \caption{Acceptance-times-efficiency factors for \Bzero mesons as a function of \pt. For each \pt interval, the acceptance correction is shown (green), then multiplied by both trigger and BDT selection efficiencies (blue).}
    \label{fig:acc_eff}
\end{figure}

The $(\mathrm{Acc} \times \varepsilon)$ correction factor was obtained from simulations, using samples distinct from those utilised in the BDT training. The $(\mathrm{Acc} \times \varepsilon)$ values computed as a function of \pt, after applying all the selections, are shown in \figref{fig:acc_eff}. The purely-geometrical acceptance, defined for this analysis by the single-track requirements of $\lvert\eta\rvert<0.8$ and $\pt>100~\mevc$, and the $\mathrm{Acc} \times \varepsilon$, including both trigger and analysis selection efficiency, are distinguished with green and blue colours, respectively. The correction factor for the rapidity coverage $c_{\Delta y}$ was computed with the MC simulations described in Section~\ref{sec:exp}, and was defined as the ratio between the generated \Bzero-meson yield in $|y|<0.8$ and that in $|y|<0.5$.

\section{Systematic uncertainties}
\label{sec:systematics}

The measurement of the \pt-differential production cross section of \Bzero mesons is affected by the following sources of systematic uncertainties: (i) raw-yield extraction from the invariant-mass distribution, (ii) \Bzero selection efficiency, (iii) single-track selection efficiency, and (iv) ITS--TPC matching efficiency. In addition, the \pt-differential production cross section is affected by an overall normalisation uncertainty due to the 3.6\% uncertainty on the branching ratio of the analysed decay channel~\cite{ParticleDataGroup:2024cfk} and the 3.9\% uncertainty on the integrated luminosity of the recorded sample. The values of the systematic uncertainties for the analysed \pt intervals are reported in \Tabref{tab:systematics}. The contributions of the different sources were considered to be uncorrelated and were summed in quadrature to obtain the total systematic uncertainty.

\begin{table}[!b]
    \centering
    \caption{Relative systematic uncertainties on the measurement of the \Bzero-meson production cross section in the analysed \pt intervals. All the reported systematic uncertainties are considered to be correlated across \pt intervals, with the exception of the raw-yield extraction ones.}
    \begin{tabular}{|c|ccccccc|}
    \hline
     \pt (\gevc)    &  1--2 & 2--4 & 4--6 & 6--8 & 8--10 & 10--14 & 14--23.5\\
     \hline
     Raw-yield extraction    & 7\% &  7\% &  7\% &  7\% &  7\% &  7\% &  15\% \\
     Selection efficiency &   14\% & 9\% & 3\% & 3\% & 3\% & 3\% & 3\% \\
     Single-track selections &  7\% & 5\% & 5\% & 5\% & 5\% & 5\% & 5\% \\
     ITS--TPC matching &  4\% & 4\% & 4\% & 4\% & 4\% & 4\% & 4\% \\
     \hline 
     Branching ratio & \multicolumn{7}{c|}{3.6\%} \\
     Luminosity & \multicolumn{7}{c|}{3.9\%} \\
     \hline
     Total & 16\% &  14\% &  11\% &  11\% &  11\% &  11\% &  17\% \\
     \hline
    \end{tabular}
    \label{tab:systematics}
\end{table}

The systematic uncertainty on the raw-yield extraction was evaluated by repeating the fits to the invariant-mass distribution for each \pt interval of the analysis varying the fit range and the functional form used to fit the combinatorial background, and avoiding fixing the ratio of the correlated background and signal yields. The systematic uncertainty was defined as the root mean square (RMS) of the distribution of the signal yields obtained from the described variations and ranges from 7\% to 15\% depending on the \pt interval.

The systematic uncertainty on the selection efficiency originates from imperfections in the description of the kinematic and topological variables of the candidates and of the detector resolutions and alignments in the simulation. It was assessed in two steps. First, the systematic uncertainty on the BDT-selection efficiency was estimated by repeating the measurement varying the BDT threshold value on the signal probability, which significantly modified the efficiency values, and by considering the RMS of the obtained corrected yields. Then, the sensitivity of the measurement to possible discrepancies in the spatial track resolution between data and MC was evaluated by comparing results obtained using simulations with the nominal resolution and with a 10\% worse resolution. Only the $1 < \pt <2 ~ \gevc$ interval was sensitive to such variations. The associated systematic uncertainty was then summed in quadrature with the BDT-selection efficiency systematic uncertainty in this particular \pt interval. The assigned systematic uncertainty ranges from  3\% to 14\% depending on \pt.

The track reconstruction efficiency is mainly driven by the track quality selections and the ITS--TPC matching efficiency. The systematic uncertainty on the former was estimated by repeating the analysis and varying the minimum number of ITS clusters, the minimum number of TPC space points, and the maximum $\chi^2/$number-of-clusters in the TPC required for the decay tracks. The assigned systematic uncertainty ranges from 5\% to 7\%. The latter was estimated via a data-driven tag-and-probe method using \DplustoKpipi and \DstartoDpi decays to evaluate the systematic uncertainty on kaon and pion track reconstruction, respectively. This method reconstructs the decay vertex of the charm hadron with two out of the three decay tracks (the decay products of the \Dzero meson in the case of the \Dstar decay). The raw signal is then extracted via an invariant-mass analysis by combining the two-track vertex with a third (probe) track. The systematic uncertainty is estimated by comparing the ratios of raw signal yields obtained with different probe track selection criteria (requiring hits only in the ITS, only in the TPC, or in both detectors) in real data and in the MC simulation. 
An uncertainty of 4\% was assigned in all \pt intervals.

\section{Results}
\label{sec:results}
\begin{figure}[!tb]
    \centering
    \includegraphics[width=1.\textwidth]{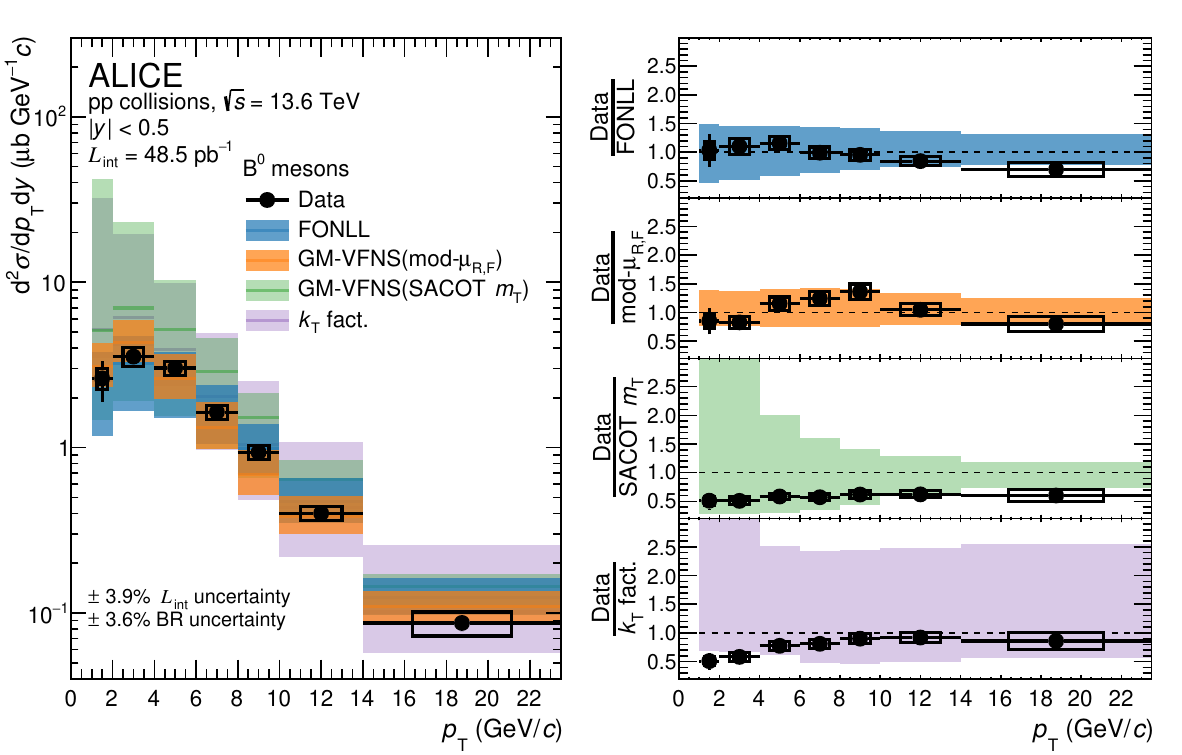}
    \caption{\pt-differential production cross section of \Bzero mesons measured at midrapidity ($|y|<0.5$) in pp collisions at $\sqrts=13.6~\TeV$ compared with FONLL~\cite{Cacciari:2012ny}, GM-VFNS(mod-\muRF)~\cite{Kniehl:2007erq,Benzke:2019usl}, GM-VFNS(SACOT-\mt)~\cite{Helenius:2023wkn}, and \kt-factorisation~\cite{Guiot:2021vnp,Barattini:2025wbo} calculations (left panel) and ratios of the data to the theoretical predictions (right panel). Statistical uncertainties are depicted as vertical error bars, and systematic uncertainties as boxes. The systematic uncertainties related to normalisation are quoted separately as text.}
    \label{fig:xsec_pqcd}
\end{figure}

The \pt-differential production cross section of \Bzero mesons in $|y|<0.5$ in pp collisions at $\sqrts=13.6~\TeV$ is shown in Fig.~\ref{fig:xsec_pqcd}. Statistical uncertainties are depicted as vertical error bars, and systematic uncertainties as boxes. The systematic uncertainties related to normalisation are quoted separately in the text. In each \pt interval, the markers are positioned horizontally in its centre and the horizontal bars represent the width of the \pt interval. The measured cross sections are compared in Appendix~\ref{app:cms_lhcb} to the \Bplus measurements reported by CMS (at midrapidity, for $\pt>10$~\gevc)~\cite{CMS:2016plw} and by the LHCb Collaboration at forward rapidities~\cite{LHCb:2017vec} for pp collisions at $\sqrts=13$~\tev.

The measurement is compared in Fig.~\ref{fig:xsec_pqcd} with pQCD calculations performed at NLO with different schemes: FONLL~\cite{Cacciari:2012ny}, GM-VFNS(mod-\muRF)~\cite{Kniehl:2007erq,Benzke:2019usl}, GM-VFNS(SACOT-\mt)~\cite{Helenius:2023wkn}, and \kt-factorisation~\cite{Guiot:2021vnp,Barattini:2025wbo}.
The FONLL uncertainty band includes (i) the uncertainties due to the choice of the renormalisation (\muR) and factorisation (\muF) scales, evaluated by varying the scale parameters by a factor of two, (ii) the uncertainties related to the value of the b-quark mass, and (iii) the uncertainties on the NNPDF3.0 PDFs~\cite{NNPDF:2014otw}.
The two GM-VFNS calculations differ from each other in the prescriptions to regulate the divergences at small transverse
momentum. In the mod-\muRF scheme, these divergences are tamed by tuning the factorisation and fragmentation scales, while in the SACOT-\mt scheme they are regulated by considering the finite heavy-quark mass, as introduced in Ref.~\cite{Helenius:2018uul}. Different sets of PDFs are used in the calculations, namely CT18 NLO~\cite{Hou:2019efy} for GM-VFNS(mod-\muRF) and NNPDF4.0~\cite{NNPDF:2021njg} for GM-VFNS(SACOT-\mt).
The \kt-factorisation calculations overcome the factorisation scheme employed in previous works (see, e.g., Ref.~\cite{Catani:1990eg}) for the estimation of the unintegrated PDFs~\cite{Guiot:2022psv}, and adopt the variable-flavour-number scheme.
Moreover, in Ref.~\cite{Barattini:2025wbo} the authors adopted for the first time scale-dependent fragmentation functions, highlighting the role of the gluon-to-heavy-hadron contribution, which improves the agreement with the data at low \pt.

All the predictions based on pQCD calculations at NLO accuracy are compatible with the measurement within uncertainties, though the theoretical uncertainties remain significantly larger than the experimental ones. The data points lie close to the central values of FONLL calculations in the full \pt range. The central values of the GM-VFNS(mod-\muRF) calculations describe the low- and high-\pt regions, while they tend to underestimate the $4<\pt<10~\GeV/c$ interval. The central values of the predictions from the GM-VFNS(SACOT-\mt) calculations overestimate the measured cross section in the full \pt interval, whereas those from the \kt-factorisation calculations, which yield larger theoretical uncertainties than the other NLO calculations, are in good agreement with the data for $\pt>4~\GeV/c$ and slightly overestimate the data at lower \pt.

\begin{figure}[!tb]
    \centering
    \includegraphics[width=0.6\linewidth]{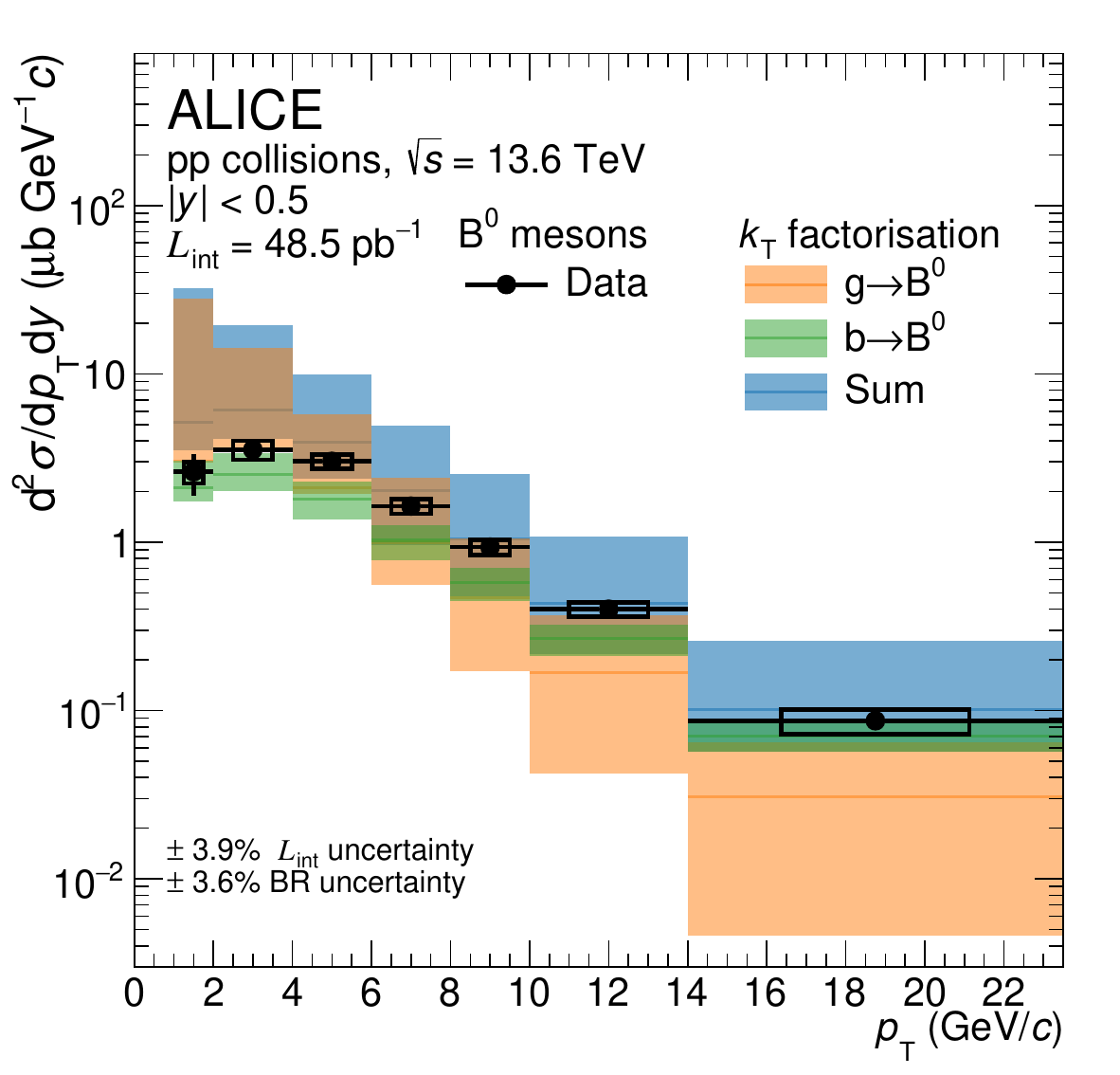}
    \caption{\pt-differential production cross section of \Bzero mesons measured at midrapidity ($|y|<0.5$) in pp collisions at $\sqrts=13.6~\TeV$ compared with \kt-factorisation~\cite{Guiot:2021vnp,Barattini:2025wbo} calculations. The contributions to the predicted \Bzero production cross section arising from gluon and beauty-quark fragmentation into a \Bzero meson are also shown. Statistical uncertainties are depicted as vertical error bars, and systematic uncertainties as boxes. The systematic uncertainties related to normalisation are quoted separately as text.}
    \label{fig:xsec_kt}
\end{figure}

When the scale evolution of the fragmentation functions is included in the \kt-factorisation calculations using the Dokshitzer--Gribov--Lipatov--Altarelli--Parisi equations~\cite{Altarelli:1977zs}, additional contributions to the production of \Bzero mesons arise from light quark- and gluon-initiated fragmentation. While the light-quark contribution is generally negligible, the gluon one can have a non-negligible effect at low \pt. Fig.~\ref{fig:xsec_kt} shows the measured \pt-differential \Bzero production cross section  compared to \kt-factorisation calculations~\cite{Guiot:2021vnp,Barattini:2025wbo} with the $\mathrm{g}\rightarrow\Bzero$ and $\mathrm{b}\rightarrow\Bzero$ contributions displayed separately. The $\mathrm{g}\rightarrow\Bzero$ process is characterised by large uncertainties at low \pt; nevertheless, the measured \Bzero production cross section can still be described using only the $\mathrm{b}\rightarrow\Bzero$ contribution in most of the considered \pt intervals.

\begin{figure}[!tb]
    \centering
    \includegraphics[width=1.\linewidth]{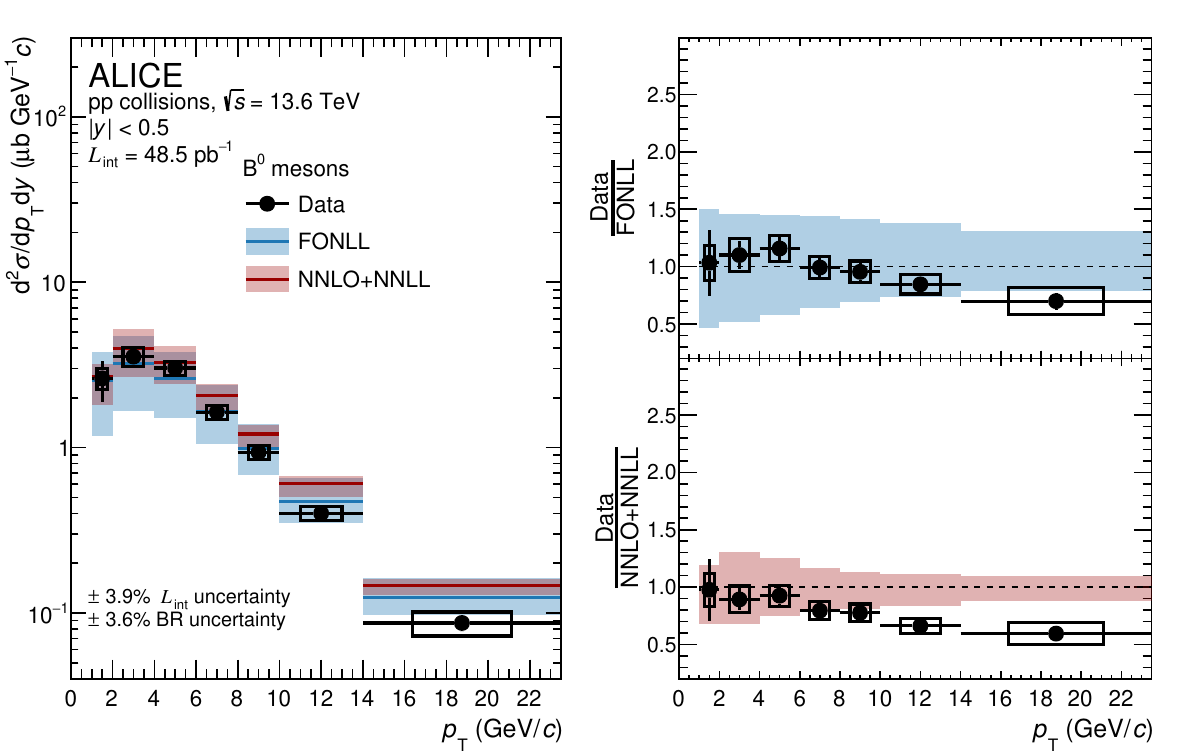}
    \caption{\pt-differential production cross section of \Bzero mesons measured at midrapidity ($|y|<0.5$) in pp collisions at $\sqrts=13.6~\TeV$ compared with FONLL~\cite{Cacciari:2012ny} and \nnlonnll~\cite{Czakon:2024tjr} calculations (left panel) and ratios of the data to the theoretical predictions (right panel). Statistical uncertainties are depicted as vertical error bars, and systematic uncertainties as boxes. The systematic uncertainties related to normalisation are quoted separately as text.}
    \label{fig:xsec_nnlo_nnll}
\end{figure}

Recent theoretical advances have enabled successful calculations at next-to-next-to-leading perturbative order with resummation of collinear logarithms at next-to-next-to-leading log (\nnlonnll) accuracy, which are characterised by a reduced dependence on energy scales~\cite{Czakon:2024tjr}. In Fig.~\ref{fig:xsec_nnlo_nnll}, the measured \pt-differential production cross section of \Bzero mesons is compared with calculations at \nnlonnll and FONLL accuracy. The two predictions are compatible within uncertainties, although \nnlonnll predictions lie on the upper part of the FONLL uncertainty band, and show a reduced uncertainty as compared to FONLL calculations. Different PDF sets are used in the two predictions: the \nnlonnll calculation employs NNPDF3.1 at NNLO accuracy~\cite{NNPDF:2017mvq}, while FONLL uses NNPDF3.0 at NLO accuracy~\cite{NNPDF:2014otw}. Similarly, the parametrisation of the fragmentation functions also differs between the two predictions. The \nnlonnll calculations use the CGMP set at NNLO perturbative precision~\cite{Czakon:2022pyz}, whereas FONLL uses a set of fragmentation functions at NLO accuracy~\cite{Cacciari:2005uk}. The \nnlonnll calculations are compatible with the measurement within the uncertainties up to $\pt < 10$~\gevc, and within about $2\sigma$ for $10<\pt<23.5$~\gevc. 

\begin{figure}[!tb]
    \centering
    \includegraphics[width=1.\textwidth]{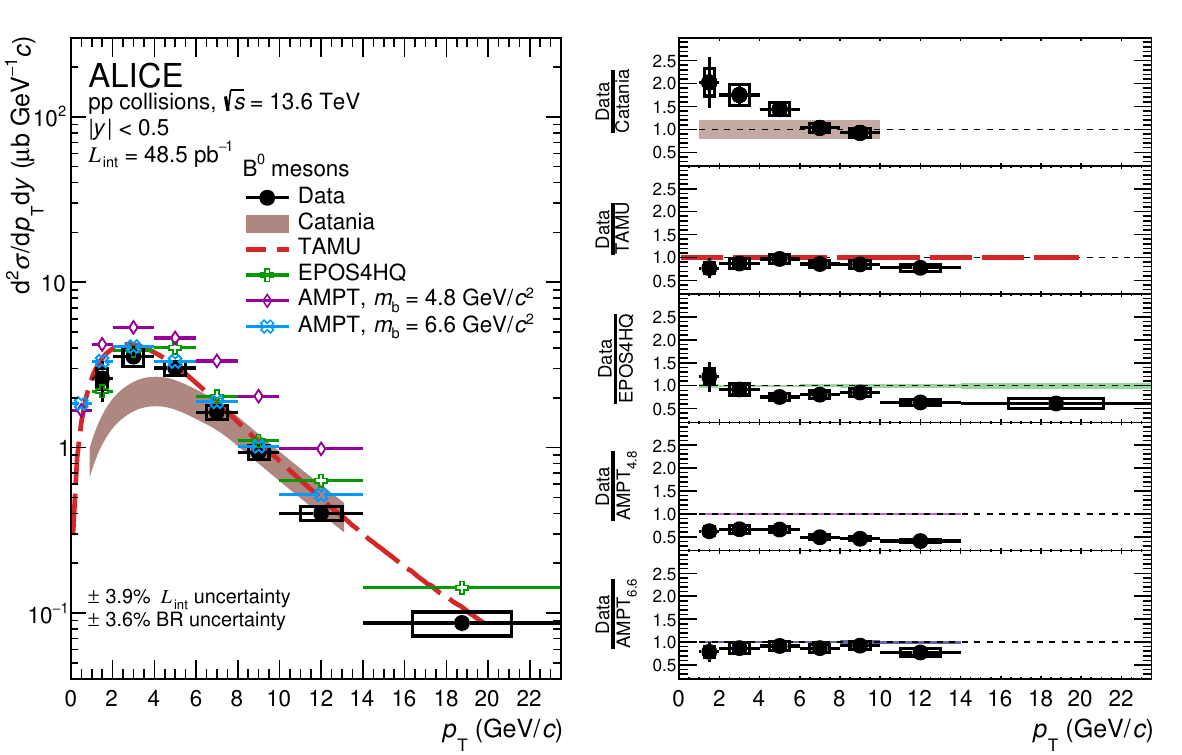}
    \caption{\pt-differential production cross section of \Bzero mesons measured at midrapidity ($|y|<0.5$) in pp collisions at $\sqrts=13.6~\TeV$ compared with Catania~\cite{Minissale:2024gxx}, TAMU~\cite{He:2022tod}, EPOS4HQ~\cite{Zhao:2024ecc, Aichelin:2024dke}, and AMPT~\cite{He:2025sfi} models (left panel) and ratios of the data to the theoretical predictions, where the latter are integrated over the width of the experimental \pt intervals. (right panel). Statistical uncertainties are depicted as vertical error bars, and systematic uncertainties as boxes. The systematic uncertainties related to normalisation are quoted separately as text.}
    \label{fig:xsec_pheno}
\end{figure}

In Fig.~\ref{fig:xsec_pheno}, the measurement is compared with pQCD-inspired phenomenological models. The TAMU model is based on the statistical hadronisation of beauty quarks~\cite{He:2022tod}. The \bbbar production cross section is obtained from LHCb measurements at forward rapidity~\cite{LHCb:2017vec} and rescaled at midrapidity with a rapidity-scaling factor from FONLL calculations, yielding  $\mathrm{d}\sigma_{\bbbar}/\mathrm{d}y|_{y=0} = 85.4~\mub$. The b-quark \pt distribution is obtained from FONLL calculations, while the branching fractions of beauty quarks to the different hadron species are assumed to follow the relative thermal densities. This model also adopts an enriched set of beauty-hadron states to be populated, expected from the RQM~\cite{Ebert:2011kk}, which includes yet-unobserved excited states. The Catania model assumes the formation of a quark--gluon plasma (QGP) also in pp collisions, and employs a combination of fragmentation and coalescence for the hadronisation of beauty quarks~\cite{Minissale:2024gxx}. In this model, both the \bbbar production cross section at midrapidity and the b-quark \pt distribution are taken from FONLL calculations. The uncertainty band corresponds to an overall normalisation uncertainty from the FONLL total \bbbar production cross section. The coalescence process is described with the Wigner formalism and the resulting coalescence probability varies with the \pt of the beauty quarks, reaching unity in the limit of $\pt \rightarrow 0$. Quarks that do not hadronise via coalescence fragment into hadrons. Within EPOS4HQ~\cite{Aichelin:2024dke,Zhao:2024ecc}, \bbbar pairs can be produced via LO and NLO processes and a QGP is formed in regions with an energy density of light partons larger than $\epsilon> 0.57~\GeV/{\rm fm}^3$. Beauty quarks finally hadronise via fragmentation or coalescence, similarly to the Catania model. In the AMPT model~\cite{He:2025sfi}, the initial conditions are generated with PYTHIA8 within the string-melting approach~\cite{Lin:2001zk}, followed by partonic interactions, hadronisation via coalescence, and hadronic rescatterings.

The data points are compatible with the TAMU model within the experimental uncertainties over the full measured \pt interval, suggesting that the fraction of beauty quarks hadronising into \Bzero mesons can be described within a statistical hadronisation approach also in small collision systems, such as pp collisions. The Catania model provides a good description of the measured \Bzero cross section for $\pt>6~\GeV/c$ and underestimates it at lower \pt, where the bulk of the production occurs and the coalescence mechanism is expected to play a dominant role~\cite{Minissale:2024gxx}. The EPOS4HQ model describes the measurement for $\pt<10~\GeV/c$ while it overestimates the data points for higher \pt. Finally, the AMPT model overestimates the data in the full \pt range when the beauty-quark mass is set to $m_\mathrm{b}=4.8~\GeV/c^2$, while it describes the measurement when a significantly higher mass, $m_\mathrm{b}=6.6~\GeV/c^2$, is used. This sensitive deviation from the beauty-quark mass value reported in the PDG~\cite{ParticleDataGroup:2024cfk} was introduced in Ref.~\cite{He:2025sfi} to effectively reduce the \bbbar production cross section which is overestimated in the PYTHIA8 MC generator, used to describe the initial conditions in the AMPT model. It is important to note that this choice might have a sizeable impact in the description of the hadronisation via coalescence, that utilises the quark mass as parameter.

\begin{figure}[!p]
    \centering
    \includegraphics[width=0.95\textwidth]{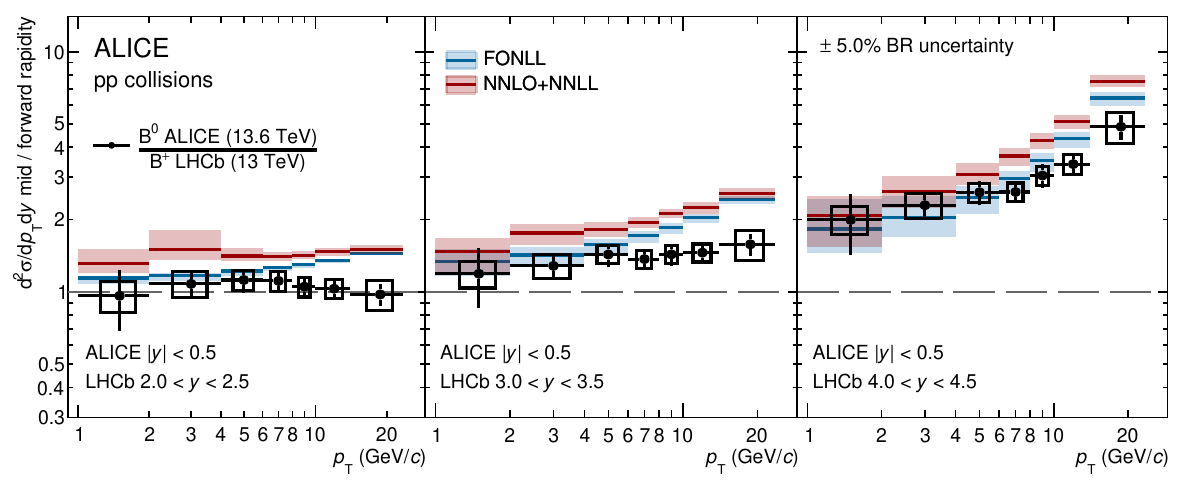}
    \includegraphics[width=0.95\textwidth]{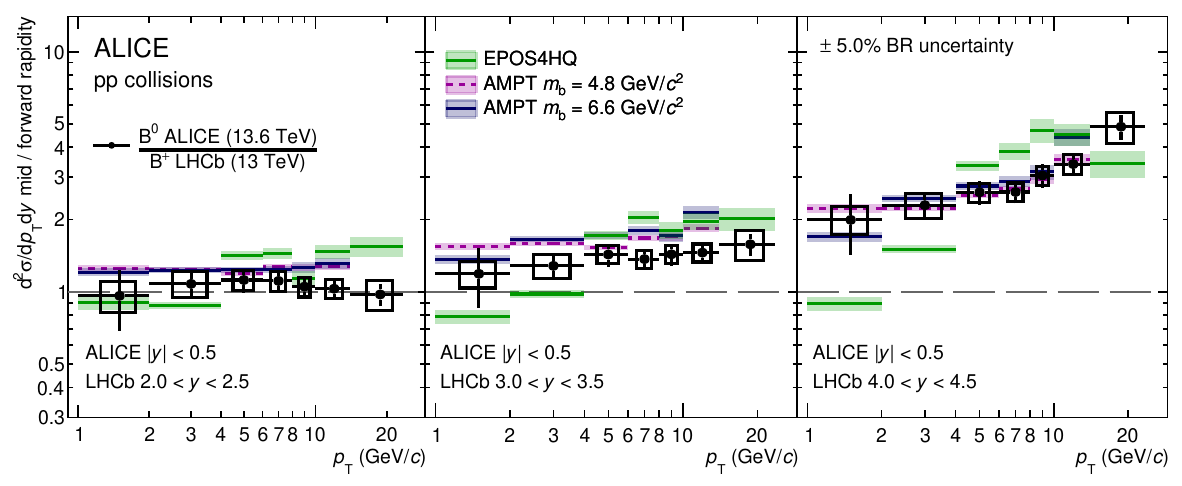}
    \includegraphics[width=0.95\textwidth]{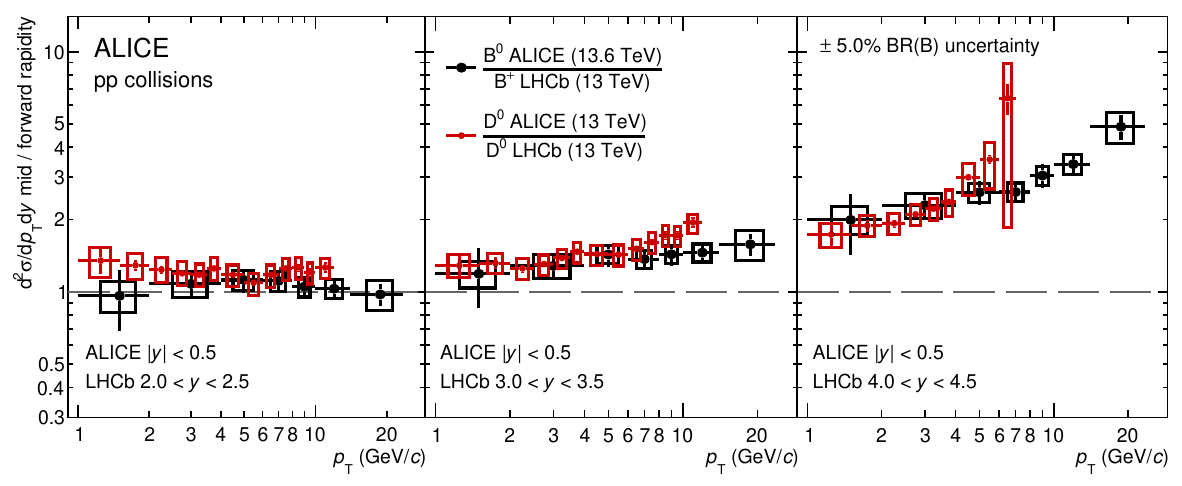}
    \caption{Ratios of \pt-differential production cross sections per unit of rapidity of \Bzero mesons at midrapidity ($|y|<0.5$) in pp collisions at $\sqrts=13.6~\TeV$ to those measured by the LHCb Collaboration~\cite{LHCb:2017vec} for \Bplus mesons in pp collisions at $\sqrts=13~\TeV$ in three intervals of rapidity, $2<y<2.5$ (left panel), $3<y<3.5$ (middle panel), and $4<y<4.5$ (right panel). Statistical uncertainties are reported as vertical error bars, and systematic uncertainties as boxes, except for BR uncertainties that are quoted separately as text. The data points are compared to pQCD calculations (FONLL~\cite{Cacciari:2012ny}, \nnlonnll~\cite{Czakon:2024tjr}) in the upper row, phenomenological models (EPOS4HQ~\cite{Aichelin:2024dke,Zhao:2024ecc} and AMPT~\cite{He:2025sfi}) in the middle row, and to the same ratio measured for \Dzero mesons in pp collisions at $\sqrts=13~\TeV$~\cite{ALICE:2023sgl} in the bottom row.}
    \label{fig:ratioLHCb}
\end{figure}

To study the rapidity dependence of B-meson production, the ratio between the \pt-differential production cross section of \Bzero mesons per unit of rapidity at midrapidity ($|y|<0.5$) in pp collisions at $\sqrts=13.6~\TeV$ was divided by the one of \Bplus mesons measured by the LHCb Collaboration in pp collisions at $\sqrts=13~\TeV$ in three different rapidity intervals, $2<y<2.5$, $3<y<3.5$, and $4<y<4.5$~\cite{LHCb:2017vec}. The statistical and systematic uncertainties, including the one on the BR of the decay channel and the luminosity determination, were treated as uncorrelated between the ALICE and LHCb results. When considering the $2<y<2.5$ rapidity interval at the denominator, the ratios do not show a significant \pt dependence, while an increase with increasing \pt is found for larger rapidities at the denominator. 
The experimental ratios are compared to FONLL and \nnlonnll calculations in the top panels of Fig.~\ref{fig:ratioLHCb}. Both calculations for the numerator and denominator were performed at the two values of $\sqrts$ corresponding to the different centre-of-mass energies of the data points. The maximum difference between the FONLL cross section at the two energies is obtained in the highest-\pT interval and is about 5\%. The predictions based on \nnlonnll calculations are systematically higher than FONLL ones. However, even if the data central values tend to lie closer to FONLL predictions, the experimental points are compatible with both calculations within about $2.5\sigma$ for all the rapidity intervals, with the larger deviations found at high \pt in the two most central rapidity intervals considered.
The measurement is also compared with the EPOS4HQ and AMPT models. The first model describes the measured ratios for $\pt>4~\GeV/c$, while it underestimates them at lower \pt. Within the measured \pt range, AMPT features a small dependency on $m_\mathrm{b}$ and is compatible with the experimental points both in the case of $m_\mathrm{b}=4.8~\GeV/c^2$ and $m_\mathrm{b}=6.6~\GeV/c^2$.

As illustrated in the bottom panels of Fig.~\ref{fig:ratioLHCb}, the ratios between the \pt-differential production cross sections measured at mid and forward rapidities for B and D mesons~\cite{ALICE:2023sgl} are compatible within uncertainties, suggesting a mild dependence of the rapidity dependency on the heavy-quark mass.

The visible production cross section of \Bzero mesons was evaluated by integrating the \pt-differential production cross section in the measured \pt range. The systematic uncertainty
was evaluated by propagating all the uncertainties as correlated among the \pt intervals, except the one related to the yield extraction, owing to the variation of the signal-to-background ratio and the background invariant-mass shape. The resulting visible cross section is 
\begin{equation}
\frac{\mathrm{d}\sigma(\Bzero)}{\dy}\bigg{|}_{|y|<0.5,~1<\pt<23.5~\GeV/c} = 23.3 \pm 1.3~(\text{stat.}) \pm 2.5~(\text{syst.})~\mub.
\end{equation}
The visible cross section was extrapolated to the production cross section per unit of rapidity using the ratio between the cross section predicted by \nnlonnll calculations for $\pt > 0$ and $|y|<0.5$ and the one in the experimentally covered phase space. This resulting extrapolation factor is $1.038^{+0.010}_{-0.007}$, where the uncertainties include the variations of the beauty-quark mass and of the factorisation and renormalisation scales, as well as the uncertainties on the PDFs and fragmentation functions. This leads to a production cross section per unit of rapidity of 
\begin{equation}
\frac{\mathrm{d}\sigma(\Bzero)}{\dy}\bigg{|}_{|y|<0.5} = 24.2 \pm 1.4~(\text{stat.}) \pm 2.6~(\text{syst.})_{-0.3}^{+0.2}~(\text{extrap.})~\mub.
\end{equation}

\section{Summary}
\label{sec:summary}

This article reports the measurement of the \pt-differential production cross section of \Bzero mesons in pp collisions at a centre-of-mass energy of $\sqrts=13.6~\TeV$ with the ALICE detector at the CERN LHC. The measurement relies on an offline trigger selection of the full minimum-bias data samples collected in 2023 and 2024. For the first time, the \Bzero production cross section is measured at midrapidity down to $\pt=1~\GeV/c$ at LHC energies, extending the kinematic range of previous results~\cite{CMS:2011pdu}. This result, after extrapolation to the relevant centre-of-mass energy~\cite{Averbeck:2011ga}, provides a crucial reference for future measurements in heavy-ion collisions, where the production of beauty hadrons is expected to exhibit modifications at low \pT due to the diffusion of beauty quarks in the colour-deconfined state of matter produced~\cite{ALICE:2022tji,ALICE:2023gjj}. It also serves as baseline for future measurements of beauty-baryon production at midrapidity in pp collisions.
The rapidity dependence of B-meson production is studied by computing the ratio of the \Bzero-meson cross section to the \Bplus-meson one measured by the LHCb Collaboration at forward rapidity. The mid-to-forward rapidity ratio of B mesons agrees with the one observed for \Dzero mesons within the experimental uncertainties. The measurements are in agreement within uncertainties with predictions based on pQCD calculations implementing different factorisation and renormalisation schemes. 
The measured production of \Bzero mesons is also compared with phenomenological models for the hadronisation. Calculations based on a statistical hadronisation approach for the fragmentation fraction of beauty quarks to \Bzero mesons along with a \bbbar \pt-differential production cross section from FONLL calculations normalised to the total \bbbar cross section from LHCb measurements describe the measurement within uncertainties, similarly to what was previously observed in the charm sector~\cite{ALICE:2023sgl}. 
Other phenomenological models implementing beauty-quark hadronisation via coalescence show discrepancies compared with the experimental results, mostly in the low-\pt region, such as the Catania model for the \pt-differential production cross section at midrapidity or the EPOS4HQ model for the \pt-differential ratios of cross sections measured in different rapidity intervals. This highlights the importance of measurements at low \pt to better constrain phenomenological models.
Finally, the total \Bzero production cross section per unit of rapidity at midrapidity $\mathrm{d}\sigma(\Bzero)/\mathrm{d} y|_{|y|<0.5} = 24.2 \pm 1.4~(\text{stat.}) \pm 2.6~(\text{syst.})_{-0.3}^{+0.2}~(\text{extrap.})~\mub$ was obtained.


\newenvironment{acknowledgement}{\relax}{\relax}
\begin{acknowledgement}
\section{Acknowledgements}

The ALICE Collaboration would like to thank all its engineers and technicians for their invaluable contributions to the construction of the experiment and the CERN accelerator teams for the outstanding performance of the LHC complex.
The ALICE Collaboration gratefully acknowledges the resources and support provided by all Grid centres and the Worldwide LHC Computing Grid (WLCG) collaboration.
The ALICE Collaboration acknowledges the following funding agencies for their support in building and running the ALICE detector:
A. I. Alikhanyan National Science Laboratory (Yerevan Physics Institute) Foundation (ANSL), State Committee of Science and World Federation of Scientists (WFS), Armenia;
Austrian Academy of Sciences, Austrian Science Fund (FWF): [M 2467-N36] and Nationalstiftung f\"{u}r Forschung, Technologie und Entwicklung, Austria;
Ministry of Communications and High Technologies, National Nuclear Research Center, Azerbaijan;
Rede Nacional de Física de Altas Energias (Renafae), Financiadora de Estudos e Projetos (Finep), Funda\c{c}\~{a}o de Amparo \`{a} Pesquisa do Estado de S\~{a}o Paulo (FAPESP) and The Sao Paulo Research Foundation  (FAPESP), Brazil;
Bulgarian Ministry of Education and Science, within the National Roadmap for Research Infrastructures 2020-2027 (object CERN), Bulgaria;
Ministry of Education of China (MOEC) , Ministry of Science \& Technology of China (MSTC) and National Natural Science Foundation of China (NSFC), China;
Ministry of Science and Education and Croatian Science Foundation, Croatia;
Centro de Aplicaciones Tecnol\'{o}gicas y Desarrollo Nuclear (CEADEN), Cubaenerg\'{\i}a, Cuba;
Ministry of Education, Youth and Sports of the Czech Republic, Czech Republic;
The Danish Council for Independent Research | Natural Sciences, the VILLUM FONDEN and Danish National Research Foundation (DNRF), Denmark;
Helsinki Institute of Physics (HIP), Finland;
Commissariat \`{a} l'Energie Atomique (CEA) and Institut National de Physique Nucl\'{e}aire et de Physique des Particules (IN2P3) and Centre National de la Recherche Scientifique (CNRS), France;
Bundesministerium f\"{u}r Forschung, Technologie und Raumfahrt (BMFTR) and GSI Helmholtzzentrum f\"{u}r Schwerionenforschung GmbH, Germany;
National Research, Development and Innovation Office, Hungary;
Department of Atomic Energy Government of India (DAE), Department of Science and Technology, Government of India (DST), University Grants Commission, Government of India (UGC) and Council of Scientific and Industrial Research (CSIR), India;
National Research and Innovation Agency - BRIN, Indonesia;
Istituto Nazionale di Fisica Nucleare (INFN), Italy;
Japanese Ministry of Education, Culture, Sports, Science and Technology (MEXT) and Japan Society for the Promotion of Science (JSPS) KAKENHI, Japan;
Consejo Nacional de Ciencia (CONACYT) y Tecnolog\'{i}a, through Fondo de Cooperaci\'{o}n Internacional en Ciencia y Tecnolog\'{i}a (FONCICYT) and Direcci\'{o}n General de Asuntos del Personal Academico (DGAPA), Mexico;
Nederlandse Organisatie voor Wetenschappelijk Onderzoek (NWO), Netherlands;
The Research Council of Norway, Norway;
Pontificia Universidad Cat\'{o}lica del Per\'{u}, Peru;
Ministry of Science and Higher Education, National Science Centre and WUT ID-UB, Poland;
Korea Institute of Science and Technology Information and National Research Foundation of Korea (NRF), Republic of Korea;
Ministry of Education and Scientific Research, Institute of Atomic Physics, Ministry of Research and Innovation and Institute of Atomic Physics and Universitatea Nationala de Stiinta si Tehnologie Politehnica Bucuresti, Romania;
Ministerstvo skolstva, vyskumu, vyvoja a mladeze SR, Slovakia;
National Research Foundation of South Africa, South Africa;
Swedish Research Council (VR) and Knut \& Alice Wallenberg Foundation (KAW), Sweden;
European Organization for Nuclear Research, Switzerland;
Suranaree University of Technology (SUT), National Science and Technology Development Agency (NSTDA) and National Science, Research and Innovation Fund (NSRF via PMU-B B05F650021), Thailand;
Turkish Energy, Nuclear and Mineral Research Agency (TENMAK), Turkey;
National Academy of  Sciences of Ukraine, Ukraine;
Science and Technology Facilities Council (STFC), United Kingdom;
National Science Foundation of the United States of America (NSF) and United States Department of Energy, Office of Nuclear Physics (DOE NP), United States of America.
In addition, individual groups or members have received support from:
FORTE project, reg.\ no.\ CZ.02.01.01/00/22\_008/0004632, Czech Republic, co-funded by the European Union, Czech Republic;
European Research Council (grant no. 950692), European Union;
Deutsche Forschungs Gemeinschaft (DFG, German Research Foundation) ``Neutrinos and Dark Matter in Astro- and Particle Physics'' (grant no. SFB 1258), Germany;
FAIR - Future Artificial Intelligence Research, funded by the NextGenerationEU program (Italy).

\end{acknowledgement}

\bibliographystyle{utphys}   
\bibliography{bibliography}

@article{CMS:2024vip,
    author = "Hayrapetyan, Aram and others",
    collaboration = "CMS",
    title = "{Bottom quark energy loss and hadronization with B$^{+}$ and $ {\textrm{B}}_{\textrm{s}}^0 $ nuclear modification factors using pp and PbPb collisions at $ \sqrt{s_{\textrm{NN}}} $ = 5.02 TeV}",
    eprint = "2409.07258",
    archivePrefix = "arXiv",
    primaryClass = "nucl-ex",
    reportNumber = "CMS-HIN-21-014, CERN-EP-2024-176",
    doi = "10.1007/JHEP02(2025)195",
    journal = "JHEP",
    volume = "02",
    pages = "195",
    year = "2025"
}

@article{ALICE:2024xln,
    author = "Acharya, Shreyasi and others",
    collaboration = "ALICE",
    title = "{Measurement of beauty-quark production in pp collisions at $\sqrt{s}$ = 13 TeV via non-prompt D mesons}",
    eprint = "2402.16417",
    archivePrefix = "arXiv",
    primaryClass = "hep-ex",
    reportNumber = "CERN-EP-2024-055",
    doi = "10.1007/JHEP10(2024)110",
    journal = "JHEP",
    volume = "10",
    pages = "110",
    year = "2024"
}

@article{ALICE:2021mgk,
    author = "Acharya, Shreyasi and others",
    collaboration = "ALICE",
    title = "{Measurement of beauty and charm production in pp collisions at $ \sqrt{s} $ = 5.02 TeV via non-prompt and prompt D mesons}",
    eprint = "2102.13601",
    archivePrefix = "arXiv",
    primaryClass = "nucl-ex",
    reportNumber = "CERN-EP-2021-034",
    doi = "10.1007/JHEP05(2021)220",
    journal = "JHEP",
    volume = "05",
    pages = "220",
    year = "2021"
}

@article{NNPDF:2017mvq,
    author = "Ball, Richard D. and others",
    collaboration = "NNPDF",
    title = "{Parton distributions from high-precision collider data}",
    eprint = "1706.00428",
    archivePrefix = "arXiv",
    primaryClass = "hep-ph",
    reportNumber = "EDINBURGH-2017-08, NIKHEF-2017-006, OUTP-17-04P, TIF-UNIMI-2017-3, CAVENDISH-HEP-17-06, CERN-TH-2017-077, Edinburgh 2017/08,
  Nikhef/2017-006, OUTP-17-04P,TIF-UNIMI-2017-3",
    doi = "10.1140/epjc/s10052-017-5199-5",
    journal = "Eur. Phys. J. C",
    volume = "77",
    number = "10",
    pages = "663",
    year = "2017"
}

@article{Czakon:2022pyz,
    author = "Czakon, Micha{\l} and Generet, Terry and Mitov, Alexander and Poncelet, Rene",
    title = "{NNLO B-fragmentation fits and their application to $ t\overline{t} $ production and decay at the LHC}",
    eprint = "2210.06078",
    archivePrefix = "arXiv",
    primaryClass = "hep-ph",
    reportNumber = "Cavendish-HEP-22/08, P3H-22-100, TTK-22-31",
    doi = "10.1007/JHEP03(2023)251",
    journal = "JHEP",
    volume = "03",
    pages = "251",
    year = "2023"
}

@article{Cacciari:2005uk,
    author = "Cacciari, Matteo and Nason, Paolo and Oleari, Carlo",
    title = "{A Study of heavy flavored meson fragmentation functions in e$^+$ e$^-$ annihilation}",
    eprint = "hep-ph/0510032",
    archivePrefix = "arXiv",
    reportNumber = "BICOCCA-FT-05-23, LPTHE-05-24",
    doi = "10.1088/1126-6708/2006/04/006",
    journal = "JHEP",
    volume = "04",
    pages = "006",
    year = "2006"
}

@article{Eschle:2019jmu,
    author = "Eschle, Jonas and Puig Navarro, Albert and Silva Coutinho, Rafael and Serra, Nicola",
    title = "{zfit: scalable pythonic fitting}",
    eprint = "1910.13429",
    archivePrefix = "arXiv",
    primaryClass = "physics.data-an",
    journal = {SoftwareX},
    volume = {11},
    pages = {100508},
    year = {2020},
    issn = {2352-7110},
    doi = {https://doi.org/10.1016/j.softx.2020.100508},
}

@article{Averbeck:2011ga,
    author = "Averbeck, R. and Bastid, N. and Conesa del Valle, Z. and Crochet, P. and Dainese, A. and Zhang, X.",
    title = "{Reference heavy flavour cross sections in pp collisions at \sqrts = 2.76 TeV, using a pQCD-driven \sqrts-scaling of ALICE measurements at \sqrts = 7 TeV}",
    eprint = "1107.3243",
    archivePrefix = "arXiv",
    primaryClass = "hep-ph",
    month = "7",
    year = "2011"
}

@article{ALICE:2021leo,
    collaboration = "ALICE",
    author = "Acharya, Shreyasi and others",
    title = "{ALICE 2016-2017-2018 luminosity determination for pp collisions at $\mathbf{\sqrt{{\textit s}}}$ = 13 TeV}",
    report = "\href{https://cds.cern.ch/record/2776672}{ALICE-PUBLIC-2021-005}",
    year = "2021"
}

@software{grosa_2023_13902597,
  author       = {Grosa, Fabrizio and
                  Politanò, Stefano and
                  Bigot, Alexandre and
                  Chinu, Fabrizio},
  title        = {flarefly},
  month        = apr,
  year         = 2023,
  publisher    = {Zenodo},
  version      = {0.0.12},
  doi          = {10.5281/zenodo.13902597},
  url          = {https://doi.org/10.5281/zenodo.13902597},
}

@article{ALICE:2023wbx,
    author = "Acharya, Shreyasi and others",
    collaboration = "ALICE",
    title = "{Study of flavor dependence of the baryon-to-meson ratio in proton--proton collisions at $\sqrt{s}=13$ TeV}",
    eprint = "2308.04873",
    archivePrefix = "arXiv",
    primaryClass = "hep-ex",
    reportNumber = "CERN-EP-2023-159",
    doi = "10.1103/PhysRevD.108.112003",
    journal = "Phys. Rev. D",
    volume = "108",
    number = "11",
    pages = "112003",
    year = "2023"
}

@article{ALICE:2021edd,
    author = "Acharya, Shreyasi and others",
    collaboration = "ALICE",
    title = "{Prompt and non-prompt J/\ensuremath{\psi} production cross sections at midrapidity in proton--proton collisions at $\sqrt{s}$ = 5.02 and 13 TeV}",
    eprint = "2108.02523",
    archivePrefix = "arXiv",
    primaryClass = "nucl-ex",
    reportNumber = "CERN-EP-2021-152",
    doi = "10.1007/JHEP03(2022)190",
    journal = "JHEP",
    volume = "03",
    pages = "190",
    year = "2022"
}

@article{ALICE:2012acz,
    author = "Abelev, Betty and others",
    collaboration = "ALICE",
    title = "{Measurement of electrons from beauty hadron decays in pp collisions at $\sqrt{s}=7$ TeV}",
    eprint = "1208.1902",
    archivePrefix = "arXiv",
    primaryClass = "hep-ex",
    reportNumber = "CERN-PH-EP-2012-229",
    doi = "10.1016/j.physletb.2013.01.069",
    journal = "Phys. Lett. B",
    volume = "721",
    pages = "13--23",
    year = "2013",
    note = "[Erratum: Phys.Lett.B 763, 507--509 (2016)]"
}

@article{ALICE:2018gev,
    author = "Acharya, Shreyasi and others",
    collaboration = "ALICE",
    title = "{Dielectron and heavy-quark production in inelastic and high-multiplicity proton--proton collisions at $\sqrt{s_{\rm NN}}=$ 13 TeV}",
    eprint = "1805.04407",
    archivePrefix = "arXiv",
    primaryClass = "hep-ex",
    reportNumber = "CERN-EP-2018-122",
    doi = "10.1016/j.physletb.2018.11.009",
    journal = "Phys. Lett. B",
    volume = "788",
    pages = "505--518",
    year = "2019"
}

@article{ALICE:2020mfy,
    author = "Acharya, Shreyasi and others",
    collaboration = "ALICE",
    title = "{Dielectron production in proton--proton and proton--lead collisions at $\sqrt{s_{\rm NN}}=$ 5.02 TeV}",
    eprint = "2005.11995",
    archivePrefix = "arXiv",
    primaryClass = "nucl-ex",
    reportNumber = "CERN-EP-2020-081",
    doi = "10.1103/PhysRevC.102.055204",
    journal = "Phys. Rev. C",
    volume = "102",
    number = "5",
    pages = "055204",
    year = "2020"
}

@article{ALICE:2023sgl,
    author = "Acharya, Shreyasi and others",
    collaboration = "ALICE",
    title = "{Charm production and fragmentation fractions at midrapidity in pp collisions at $ \sqrt{\textrm{s}} $ = 13 TeV}",
    eprint = "2308.04877",
    archivePrefix = "arXiv",
    primaryClass = "hep-ex",
    reportNumber = "CERN-EP-2023-162",
    doi = "10.1007/JHEP12(2023)086",
    journal = "JHEP",
    volume = "12",
    pages = "086",
    year = "2023"
}

@article{CMS:2012pgw,
    author = "Chatrchyan, Serguei and others",
    collaboration = "CMS",
    title = "{Inclusive $b$-Jet Production in $pp$ Collisions at $\sqrt{s}=7$ TeV}",
    eprint = "1202.4617",
    archivePrefix = "arXiv",
    primaryClass = "hep-ex",
    reportNumber = "CMS-BPH-11-022, CERN-PH-EP-2012-036",
    doi = "10.1007/JHEP04(2012)084",
    journal = "JHEP",
    volume = "04",
    pages = "084",
    year = "2012"
}

@article{CMS:2011xhf,
    author = "Khachatryan, Vardan and others",
    collaboration = "CMS",
    title = "{Inclusive b-Hadron Production Cross Section with Muons in $pp$ Collisions at $\sqrt{s} = 7$ TeV}",
    eprint = "1101.3512",
    archivePrefix = "arXiv",
    primaryClass = "hep-ex",
    reportNumber = "CERN-PH-EP-2010-088, CMS-BPH-10-007",
    doi = "10.1007/JHEP03(2011)090",
    journal = "JHEP",
    volume = "03",
    pages = "090",
    year = "2011"
}

@article{CMS:2011pdu,
    author = "Chatrchyan, Serguei and others",
    collaboration = "CMS",
    title = "{Measurement of the $B^0$ production cross section in $pp$ Collisions at $\sqrt{s}=7$ TeV}",
    eprint = "1104.2892",
    archivePrefix = "arXiv",
    primaryClass = "hep-ex",
    reportNumber = "CERN-PH-EP-2011-034, CMS-BPH-10-005",
    doi = "10.1103/PhysRevLett.106.252001",
    journal = "Phys. Rev. Lett.",
    volume = "106",
    pages = "252001",
    year = "2011"
}

@article{CMS:2016plw,
    author = "Khachatryan, Vardan and others",
    collaboration = "CMS",
    title = "{Measurement of the total and differential inclusive $B^+$ hadron cross sections in pp collisions at $\sqrt{s}$ = 13 TeV}",
    eprint = "1609.00873",
    archivePrefix = "arXiv",
    primaryClass = "hep-ex",
    reportNumber = "CMS-BPH-15-004, CERN-EP-2016-198",
    doi = "10.1016/j.physletb.2017.05.074",
    journal = "Phys. Lett. B",
    volume = "771",
    pages = "435--456",
    year = "2017"
}

@article{ATLAS:2015zdw,
    author = "Aad, Georges and others",
    collaboration = "ATLAS",
    title = "{Measurement of the differential cross-sections of prompt and non-prompt production of $J/\psi $ and $\psi (2\mathrm {S})$ in $pp$ collisions at $\sqrt{s} = 7$ and 8 TeV with the ATLAS detector}",
    eprint = "1512.03657",
    archivePrefix = "arXiv",
    primaryClass = "hep-ex",
    reportNumber = "CERN-PH-EP-2015-292",
    doi = "10.1140/epjc/s10052-016-4050-8",
    journal = "Eur. Phys. J. C",
    volume = "76",
    number = "5",
    pages = "283",
    year = "2016"
}

@article{ATLAS:2011ac,
    author = "Aad, Georges and others",
    collaboration = "ATLAS",
    title = "{Measurement of the inclusive and dijet cross-sections of $b^-$ jets in $pp$ collisions at $\sqrt{s}=7$ TeV with the ATLAS detector}",
    eprint = "1109.6833",
    archivePrefix = "arXiv",
    primaryClass = "hep-ex",
    reportNumber = "CERN-PH-EP-2011-146",
    doi = "10.1140/epjc/s10052-011-1846-4",
    journal = "Eur. Phys. J. C",
    volume = "71",
    pages = "1846",
    year = "2011"
}

@article{ATLAS:2012sfc,
    author = "Aad, Georges and others",
    collaboration = "ATLAS",
    title = "{Measurement of the b-hadron production cross section using decays to $D^{*}\mu^-X$ final states in pp collisions at $\sqrt{s} = 7$~TeV with the ATLAS detector}",
    eprint = "1206.3122",
    archivePrefix = "arXiv",
    primaryClass = "hep-ex",
    reportNumber = "CERN-PH-EP-2012-121",
    doi = "10.1016/j.nuclphysb.2012.07.009",
    journal = "Nucl. Phys. B",
    volume = "864",
    pages = "341--381",
    year = "2012"
}

@article{ATLAS:2013cia,
    author = "Aad, Georges and others",
    collaboration = "ATLAS",
    title = "{Measurement of the differential cross-section of $B^{+}$ meson production in pp collisions at $\sqrt{s}$ = 7 TeV at ATLAS}",
    eprint = "1307.0126",
    archivePrefix = "arXiv",
    primaryClass = "hep-ex",
    reportNumber = "CERN-PH-EP-2013-089",
    doi = "10.1007/JHEP10(2013)042",
    journal = "JHEP",
    volume = "10",
    pages = "042",
    year = "2013"
}

@article{LHCb:2015foc,
    author = "Aaij, Roel and others",
    collaboration = "LHCb",
    title = "{Measurement of forward $J/\psi$ production cross-sections in $pp$ collisions at $\sqrt{s}=13$ TeV}",
    eprint = "1509.00771",
    archivePrefix = "arXiv",
    primaryClass = "hep-ex",
    reportNumber = "LHCB-PAPER-2015-037, CERN-PH-EP-2015-222",
    doi = "10.1007/JHEP10(2015)172",
    journal = "JHEP",
    volume = "10",
    pages = "172",
    year = "2015",
    note = "[Erratum: JHEP 05, 063 (2017)]"
}

@article{LHCb:2017vec,
    author = "Aaij, Roel and others",
    collaboration = "LHCb",
    title = "{Measurement of the $B^{\pm}$ production cross-section in pp collisions at $\sqrt{s} =$ 7 and 13 TeV}",
    eprint = "1710.04921",
    archivePrefix = "arXiv",
    primaryClass = "hep-ex",
    reportNumber = "CERN-EP-2017-254, LHCB-PAPER-2017-037",
    doi = "10.1007/JHEP12(2017)026",
    journal = "JHEP",
    volume = "12",
    pages = "026",
    year = "2017"
}

@article{LHCb:2019fns,
    author = "Aaij, Roel and others",
    collaboration = "LHCb",
    title = "{Measurement of $b$ hadron fractions in 13 TeV $pp$ collisions}",
    eprint = "1902.06794",
    archivePrefix = "arXiv",
    primaryClass = "hep-ex",
    reportNumber = "CERN-EP-2019-016, LHCb-PAPER-2018-050",
    doi = "10.1103/PhysRevD.100.031102",
    journal = "Phys. Rev. D",
    volume = "100",
    number = "3",
    pages = "031102",
    year = "2019"
}

@article{LHCb:2023wbo,
    author = "Aaij, Roel and others",
    collaboration = "LHCb",
    title = "{Enhanced Production of $\Lambda_b^0$ Baryons in High-Multiplicity pp Collisions at $\sqrt{s}=13$ TeV}",
    eprint = "2310.12278",
    archivePrefix = "arXiv",
    primaryClass = "hep-ex",
    reportNumber = "LHCb-PAPER-2023-027, CERN-EP-2023-208",
    doi = "10.1103/PhysRevLett.132.081901",
    journal = "Phys. Rev. Lett.",
    volume = "132",
    number = "8",
    pages = "081901",
    year = "2024"
}

@article{LHCb:2016qpe,
    author = "Aaij, Roel and others",
    collaboration = "LHCb",
    title = "{Measurement of the $b$-quark production cross-section in 7 and 13 TeV $pp$ collisions}",
    eprint = "1612.05140",
    archivePrefix = "arXiv",
    primaryClass = "hep-ex",
    reportNumber = "CERN-EP-2016-201, LHCB-PAPER-2016-031",
    doi = "10.1103/PhysRevLett.118.052002",
    journal = "Phys. Rev. Lett.",
    volume = "118",
    number = "5",
    pages = "052002",
    year = "2017",
    note = "[Erratum: Phys.Rev.Lett. 119, 169901 (2017)]"
}

@article{UA1:1988iqx,
    author = "Albajar, C. and others",
    editor = "Tran Thanh Van, J.",
    collaboration = "UA1",
    title = "{Measurement of the Bottom Quark Production Cross-Section in Proton-anti-Proton Collisions at $\sqrt{s}$ = 0.63 TeV}",
    reportNumber = "CERN-EP-88-100",
    doi = "10.1016/0370-2693(88)91785-6",
    journal = "Phys. Lett. B",
    volume = "213",
    pages = "405",
    year = "1988"
}

@article{UA1:1990vvp,
    author = "Albajar, C. and others",
    collaboration = "UA1",
    title = "{Beauty production at the CERN p anti-p collider}",
    reportNumber = "CERN-PPE-90-155-REV, CERN-PPE-90-155",
    doi = "10.1016/0370-2693(91)90228-I",
    journal = "Phys. Lett. B",
    volume = "256",
    pages = "121--128",
    year = "1991",
    note = "[Erratum: Phys.Lett.B 262, 497 (1991)]"
}

@article{CDF:2004jtw,
    author = "Acosta, D. and others",
    collaboration = "CDF",
    title = "{Measurement of the $J/\psi$ meson and $b-$hadron production cross sections in $p\overline{p}$ collisions at $\sqrt{s} = 1960$ GeV}",
    eprint = "hep-ex/0412071",
    archivePrefix = "arXiv",
    reportNumber = "FERMILAB-PUB-04-440-E",
    doi = "10.1103/PhysRevD.71.032001",
    journal = "Phys. Rev. D",
    volume = "71",
    pages = "032001",
    year = "2005"
}

@article{CDF:2006ipg,
    author = "Abulencia, A. and others",
    collaboration = "CDF",
    title = "{Measurement of the $B^+$ production cross-section in $p\overline{p}$ collisions at $\sqrt{s}$ = 1960 GeV}",
    eprint = "hep-ex/0612015",
    archivePrefix = "arXiv",
    reportNumber = "FERMILAB-PUB-06-458-E",
    doi = "10.1103/PhysRevD.75.012010",
    journal = "Phys. Rev. D",
    volume = "75",
    pages = "012010",
    year = "2007"
}

@article{CDF:2009bqp,
    author = "Aaltonen, T. and others",
    collaboration = "CDF",
    title = "{Measurement of the b-Hadron Production Cross Section Using Decays to $\mu^-D^0X$ Final States in $p\overline{p}$ Collisions at $\sqrt{s}$ = 1.96 TeV}",
    eprint = "0903.2403",
    archivePrefix = "arXiv",
    primaryClass = "hep-ex",
    reportNumber = "FERMILAB-PUB-09-069-E",
    doi = "10.1103/PhysRevD.79.092003",
    journal = "Phys. Rev. D",
    volume = "79",
    pages = "092003",
    year = "2009"
}

@article{D0:1994vpd,
    author = "Abachi, S. and others",
    collaboration = "D0",
    title = "{Inclusive $\mu$ and $B$ quark production cross-sections in $p\bar{p}$ collisions at $\sqrt{s} = 1.8$ TeV}",
    reportNumber = "FERMILAB-PUB-94-409-E",
    doi = "10.1103/PhysRevLett.74.3548",
    journal = "Phys. Rev. Lett.",
    volume = "74",
    pages = "3548--3552",
    year = "1995"
}

@article{STAR:2010ibw,
    author = "Aggarwal, M. M. and others",
    collaboration = "STAR",
    title = "{Measurement of the Bottom contribution to non-photonic electron production in $p+p$ collisions at $\sqrt{s} $=200 GeV}",
    eprint = "1007.1200",
    archivePrefix = "arXiv",
    primaryClass = "nucl-ex",
    doi = "10.1103/PhysRevLett.105.202301",
    journal = "Phys. Rev. Lett.",
    volume = "105",
    pages = "202301",
    year = "2010"
}

@article{PHENIX:2019pxh,
    author = "Aidala, C. and others",
    collaboration = "PHENIX",
    title = "{Measurement of charm and bottom production from semileptonic hadron decays in $p+p$ collisions at $\sqrt{s_{NN}}=200$ GeV}",
    eprint = "1901.08405",
    archivePrefix = "arXiv",
    primaryClass = "hep-ex",
    doi = "10.1103/PhysRevD.99.092003",
    journal = "Phys. Rev. D",
    volume = "99",
    number = "9",
    pages = "092003",
    year = "2019"
}

@article{Collins:1989gx,
    author = "Collins, John C. and Soper, Davison E. and Sterman, George F.",
    title = "{Factorization of Hard Processes in QCD}",
    eprint = "hep-ph/0409313",
    archivePrefix = "arXiv",
    reportNumber = "ITP-SB-89-31",
    doi = "10.1142/9789814503266_0001",
    journal = "Adv. Ser. Direct. High Energy Phys.",
    volume = "5",
    pages = "1--91",
    year = "1989"
}

@article{Minissale:2020bif,
    author = "Minissale, Vincenzo and Plumari, Salvatore and Greco, Vincenzo",
    title = "{Charm hadrons in pp collisions at LHC energy within a coalescence plus fragmentation approach}",
    eprint = "2012.12001",
    archivePrefix = "arXiv",
    primaryClass = "hep-ph",
    doi = "10.1016/j.physletb.2021.136622",
    journal = "Phys. Lett. B",
    volume = "821",
    pages = "136622",
    year = "2021"
}

@article{Beraudo:2023nlq,
    author = "Beraudo, Andrea and De Pace, Arturo and Pablos, Daniel and Prino, Francesco and Monteno, Marco and Nardi, Marzia",
    title = "{Heavy-flavor transport and hadronization in pp collisions}",
    eprint = "2306.02152",
    archivePrefix = "arXiv",
    primaryClass = "hep-ph",
    doi = "10.1103/PhysRevD.109.L011501",
    journal = "Phys. Rev. D",
    volume = "109",
    number = "1",
    pages = "L011501",
    year = "2024"
}

@article{He:2019tik,
    author = "He, Min and Rapp, Ralf",
    title = "{Charm-Baryon Production in Proton-Proton Collisions}",
    eprint = "1902.08889",
    archivePrefix = "arXiv",
    primaryClass = "nucl-th",
    doi = "10.1016/j.physletb.2019.06.004",
    journal = "Phys. Lett. B",
    volume = "795",
    pages = "117--121",
    year = "2019"
}

@article{He:2022tod,
    author = "He, Min and Rapp, Ralf",
    title = "{Bottom Hadrochemistry in High-Energy Hadronic Collisions}",
    eprint = "2209.13419",
    archivePrefix = "arXiv",
    primaryClass = "hep-ph",
    doi = "10.1103/PhysRevLett.131.012301",
    journal = "Phys. Rev. Lett.",
    volume = "131",
    number = "1",
    pages = "012301",
    year = "2023"
}

@article{Altmann:2024kwx,
    author = "Altmann, J. and Dubla, A. and Greco, V. and Rossi, A. and Skands, P.",
    title = "{Towards the understanding of heavy quarks hadronization: from leptonic to heavy-ion collisions}",
    eprint = "2405.19137",
    archivePrefix = "arXiv",
    primaryClass = "hep-ph",
    doi = "10.1140/epjc/s10052-024-13641-5",
    journal = "Eur. Phys. J. C",
    volume = "85",
    number = "1",
    pages = "16",
    year = "2025"
}

@article{Minissale:2024gxx,
    author = "Minissale, Vincenzo and Greco, Vincenzo and Plumari, Salvatore",
    title = "{Bottomed mesons and baryons production in pp collisions at $\sqrt{s}=5$~TeV LHC energy within a Coalescence plus Fragmentation approach}",
    eprint = "2405.19244",
    archivePrefix = "arXiv",
    primaryClass = "hep-ph",
    doi = "10.1016/j.physletb.2024.139190",
    journal = "Phys. Lett. B",
    volume = "860",
    pages = "139190",
    year = "2025"
}

@article{Cacciari:2012ny,
    author = "Cacciari, Matteo and Frixione, Stefano and Houdeau, Nicolas and Mangano, Michelangelo L. and Nason, Paolo and Ridolfi, Giovanni",
    title = "{Theoretical predictions for charm and bottom production at the LHC}",
    eprint = "1205.6344",
    archivePrefix = "arXiv",
    primaryClass = "hep-ph",
    reportNumber = "CERN-PH-TH-2011-227",
    doi = "10.1007/JHEP10(2012)137",
    journal = "JHEP",
    volume = "10",
    pages = "137",
    year = "2012"
}

@article{Helenius:2018uul,
    author = "Helenius, Ilkka and Paukkunen, Hannu",
    title = "{Revisiting the D-meson hadroproduction in general-mass variable flavour number scheme}",
    eprint = "1804.03557",
    archivePrefix = "arXiv",
    primaryClass = "hep-ph",
    doi = "10.1007/JHEP05(2018)196",
    journal = "JHEP",
    volume = "05",
    pages = "196",
    year = "2018"
}

@article{NNPDF:2021njg,
    author = "Ball, Richard D. and others",
    collaboration = "NNPDF",
    title = "{The path to proton structure at 1\% accuracy}",
    eprint = "2109.02653",
    archivePrefix = "arXiv",
    primaryClass = "hep-ph",
    reportNumber = "Edinburgh 2021/12, Nikhef-2021-013, TIF-UNIMI-2021-11",
    doi = "10.1140/epjc/s10052-022-10328-7",
    journal = "Eur. Phys. J. C",
    volume = "82",
    number = "5",
    pages = "428",
    year = "2022"
}

@article{NNPDF:2014otw,
    author = "Ball, Richard D. and others",
    collaboration = "NNPDF",
    title = "{Parton distributions for the LHC Run II}",
    eprint = "1410.8849",
    archivePrefix = "arXiv",
    primaryClass = "hep-ph",
    reportNumber = "EDINBURGH-2014-15, IFUM-1034-FT, CERN-PH-TH-2013-253, OUTP-14-11P, CAVENDISH-HEP-14-11",
    doi = "10.1007/JHEP04(2015)040",
    journal = "JHEP",
    volume = "04",
    pages = "040",
    year = "2015"
}

@article{Catani:1990eg,
    author = "Catani, S. and Ciafaloni, M. and Hautmann, F.",
    title = "{High-energy factorization and small x heavy flavor production}",
    reportNumber = "CAVENDISH-HEP-90-27",
    doi = "10.1016/0550-3213(91)90055-3",
    journal = "Nucl. Phys. B",
    volume = "366",
    pages = "135--188",
    year = "1991"
}

@article{Shabelski:2004qy,
    author = "Shabelski, Yu. M. and Shuvaev, A. G.",
    title = "{Heavy quark hadroproduction in $k_{T}$ -factorization approach with unintegrated gluon distributions}",
    eprint = "hep-ph/0406157",
    archivePrefix = "arXiv",
    doi = "10.1134/S1063778806020165",
    journal = "Phys. Atom. Nucl.",
    volume = "69",
    pages = "314--327",
    year = "2006"
}

@article{Shabelski:2017kmy,
    author = "Shabelski, Yu. M. and Shuvaev, A. G. and Surnin, I. V.",
    title = "{Heavy quark production in $k_t$ factorization approach at LHC energies}",
    eprint = "1709.02203",
    archivePrefix = "arXiv",
    primaryClass = "hep-ph",
    doi = "10.1142/S0217751X18500033",
    journal = "Int. J. Mod. Phys. A",
    volume = "33",
    number = "01",
    pages = "1850003",
    year = "2017"
}

@article{Maciula:2019izq,
    author = "Maciula, Rafal and Szczurek, Antoni",
    title = "{Consistent treatment of charm production in higher-orders at tree-level within $k_T$-factorization approach}",
    eprint = "1905.06697",
    archivePrefix = "arXiv",
    primaryClass = "hep-ph",
    doi = "10.1103/PhysRevD.100.054001",
    journal = "Phys. Rev. D",
    volume = "100",
    number = "5",
    pages = "054001",
    year = "2019"
}

@article{Ebert:2011kk,
    author = "Ebert, D. and Faustov, R. N. and Galkin, V. O.",
    title = "{Spectroscopy and Regge trajectories of heavy baryons in the relativistic quark-diquark picture}",
    eprint = "1105.0583",
    archivePrefix = "arXiv",
    primaryClass = "hep-ph",
    reportNumber = "HU-EP-11-21",
    doi = "10.1103/PhysRevD.84.014025",
    journal = "Phys. Rev. D",
    volume = "84",
    pages = "014025",
    year = "2011"
}

@article{Kniehl:2007erq,
    author = "Kniehl, Bernd A. and Kramer, Gustav and Schienbein, Ingo and Spiesberger, Hubert",
    title = "{Finite-mass effects on inclusive $B$ meson hadroproduction}",
    eprint = "0705.4392",
    archivePrefix = "arXiv",
    primaryClass = "hep-ph",
    reportNumber = "DESY-07-066, MZ-TH-07-07, LPSC-07-46",
    doi = "10.1103/PhysRevD.77.014011",
    journal = "Phys. Rev. D",
    volume = "77",
    pages = "014011",
    year = "2008"
}

@article{Benzke:2019usl,
    author = "Benzke, M. and Kniehl, B. A. and Kramer, G. and Schienbein, I. and Spiesberger, H.",
    title = "{B-meson production in the general-mass variable-flavour-number scheme and LHC data}",
    eprint = "1907.12456",
    archivePrefix = "arXiv",
    primaryClass = "hep-ph",
    reportNumber = "DESY-19-115, DESY 19-115, MITP/19-045",
    doi = "10.1140/epjc/s10052-019-7326-y",
    journal = "Eur. Phys. J. C",
    volume = "79",
    number = "10",
    pages = "814",
    year = "2019"
}

@article{Helenius:2023wkn,
    author = "Helenius, Ilkka and Paukkunen, Hannu",
    title = "{B-meson hadroproduction in the SACOT-m$_{T}$ scheme}",
    eprint = "2303.17864",
    archivePrefix = "arXiv",
    primaryClass = "hep-ph",
    doi = "10.1007/JHEP07(2023)054",
    journal = "JHEP",
    volume = "07",
    pages = "054",
    year = "2023"
}

@article{Guiot:2021vnp,
    author = "Guiot, B. and van Hameren, A.",
    title = "{D and B-meson production using kt-factorization calculations in a variable-flavor-number scheme}",
    eprint = "2108.06419",
    archivePrefix = "arXiv",
    primaryClass = "hep-ph",
    reportNumber = "IFJPAN-IV-2021-13",
    doi = "10.1103/PhysRevD.104.094038",
    journal = "Phys. Rev. D",
    volume = "104",
    number = "9",
    pages = "094038",
    year = "2021"
}

@article{Barattini:2025wbo,
    author = "Barattini, F. E. and Dib, C. O. and Guiot, B.",
    title = "{Heavy-hadron production based on k$_{t}$-factorization with scale-dependent fragmentation functions}",
    eprint = "2501.17662",
    archivePrefix = "arXiv",
    primaryClass = "hep-ph",
    doi = "10.1007/JHEP05(2025)115",
    journal = "JHEP",
    volume = "05",
    pages = "115",
    year = "2025"
}

@article{Altarelli:1977zs,
    author = "Altarelli, Guido and Parisi, G.",
    title = "{Asymptotic Freedom in Parton Language}",
    reportNumber = "LPTENS-77-6",
    doi = "10.1016/0550-3213(77)90384-4",
    journal = "Nucl. Phys. B",
    volume = "126",
    pages = "298--318",
    year = "1977"
}

@article{Catani:2020kkl,
    author = "Catani, Stefano and Devoto, Simone and Grazzini, Massimiliano and Kallweit, Stefan and Mazzitelli, Javier",
    title = "{Bottom-quark production at hadron colliders: fully differential predictions in NNLO QCD}",
    eprint = "2010.11906",
    archivePrefix = "arXiv",
    primaryClass = "hep-ph",
    reportNumber = "ZU-TH 36/20, TIF-UNIMI-2020-27, MPP-2020-186",
    doi = "10.1007/JHEP03(2021)029",
    journal = "JHEP",
    volume = "03",
    pages = "029",
    year = "2021"
}

@article{Mazzitelli:2023znt,
    author = "Mazzitelli, Javier and Ratti, Alessandro and Wiesemann, Marius and Zanderighi, Giulia",
    title = "{B-hadron production at the LHC from bottom-quark pair production at NNLO+PS}",
    eprint = "2302.01645",
    archivePrefix = "arXiv",
    primaryClass = "hep-ph",
    doi = "10.1016/j.physletb.2023.137991",
    journal = "Phys. Lett. B",
    volume = "843",
    pages = "137991",
    year = "2023"
}

@article{Czakon:2024tjr,
    author = "Czakon, Micha{\l} and Generet, Terry and Mitov, Alexander and Poncelet, Rene",
    title = "{Open B-Hadron Production at Hadron Colliders in QCD at Next-to-Next-to-Leading-Order and Next-to-Next-to-Leading-Logarithmic Accuracy}",
    eprint = "2411.09684",
    archivePrefix = "arXiv",
    primaryClass = "hep-ph",
    reportNumber = "Cavendish-HEP-24/01, IFJPAN-IV-2024-13, P3H-24-087, TTK-24-48",
    doi = "10.1103/b6pf-rj4h",
    journal = "Phys. Rev. Lett.",
    volume = "135",
    number = "16",
    pages = "161903",
    year = "2025"
}

@software{hipe4ml,
  author       = {Barioglio, Luca and
                  Catalano, Fabio and
                  Concas, Matteo and
                  Fecchio, Pietro and
                  Grosa, Fabrizio and
                  Mazzaschi, Francesco and
                  Puccio, Maximiliano},
  title        = {hipe4ml/hipe4ml},
  month        = jul,
  year         = 2021,
  publisher    = {Zenodo},
  version      = {v0.0.14},
  doi          = {10.5281/zenodo.7014886},
  url          = {https://doi.org/10.5281/zenodo.7014886}
}

@article{ParticleDataGroup:2024cfk,
    author = "Navas, S. and others",
    collaboration = "Particle Data Group",
    title = "{Review of particle physics}",
    doi = "10.1103/PhysRevD.110.030001",
    journal = "Phys. Rev. D",
    volume = "110",
    number = "3",
    pages = "030001",
    year = "2024"
}

@article{ALICE:2014sbx,
    author = "Abelev, Betty Bezverkhny and others",
    collaboration = "ALICE",
    title = "{Performance of the ALICE Experiment at the CERN LHC}",
    eprint = "1402.4476",
    archivePrefix = "arXiv",
    primaryClass = "nucl-ex",
    reportNumber = "CERN-PH-EP-2014-031",
    doi = "10.1142/S0217751X14300440",
    journal = "Int. J. Mod. Phys. A",
    volume = "29",
    pages = "1430044",
    year = "2014"
}

@article{Parzen:Kde,
author = {Emanuel Parzen},
title = {{On Estimation of a Probability Density Function and Mode}},
volume = {33},
journal = {The Annals of Mathematical Statistics},
number = {3},
publisher = {Institute of Mathematical Statistics},
pages = {1065 -- 1076},
year = {1962},
doi = {10.1214/aoms/1177704472},
}

@article{Rosenblatt:Kde,
author = {Murray Rosenblatt},
title = {{Remarks on Some Nonparametric Estimates of a Density Function}},
volume = {27},
journal = {The Annals of Mathematical Statistics},
number = {3},
publisher = {Institute of Mathematical Statistics},
pages = {832 -- 837},
year = {1956},
doi = {10.1214/aoms/1177728190},
}

@article{Guiot:2022psv,
    author = "Guiot, Benjamin",
    title = "{Normalization of unintegrated parton densities}",
    eprint = "2205.02873",
    archivePrefix = "arXiv",
    primaryClass = "hep-ph",
    doi = "10.1103/PhysRevD.107.014015",
    journal = "Phys. Rev. D",
    volume = "107",
    number = "1",
    pages = "014015",
    year = "2023",
    note = "[Erratum: Phys.Rev.D 110, 039901 (2024)]"
}

@article{ALICE-PUBLIC-2020-005,
      collaboration = "ALICE",
      author = "Acharya, Shreyasi and others",
      title         = "{Future high-energy pp programme with ALICE}",
      year          = "2020",
      report  = "\href{https://cds.cern.ch/record/2724925}{ALICE-PUBLIC-2020-005, CERN-LHCC-2020-018, LHCC-G-179}",
      journal       = "ALICE-PUBLIC-2020-005"
}

@article{ALICE:2023udb,
    author = "Acharya, Shreyasi and others",
    collaboration = "ALICE",
    title = "{ALICE upgrades during the LHC Long Shutdown 2}",
    eprint = "2302.01238",
    archivePrefix = "arXiv",
    primaryClass = "physics.ins-det",
    reportNumber = "CERN-EP-2023-009",
    doi = "10.1088/1748-0221/19/05/P05062",
    journal = "JINST",
    volume = "19",
    number = "05",
    pages = "P05062",
    year = "2024"
}

@article{ALICE:2013nwm,
    author = "Abelev, B and others",
    collaboration = "ALICE",
    title = "{Technical Design Report for the Upgrade of the ALICE Inner Tracking System}",
    report = "\href{https://cds.cern.ch/record/1625842?ln=it}{CERN-LHCC-2013-024, ALICE-TDR-017}",
    doi = "10.1088/0954-3899/41/8/087002",
    journal = "J. Phys. G",
    volume = "41",
    pages = "087002",
    year = "2014"
}

@article{Aichelin:2024dke,
    author = "Aichelin, Joerg and Zhao, J. and Gossiaux, P. B. and Werner, K.",
    title = "{Beauty hadron production in high energy proton-proton and heavy-ion collisions in the EPOS4HQ framework}",
    eprint = "2410.15114",
    archivePrefix = "arXiv",
    primaryClass = "hep-ph",
    doi = "10.22323/1.465.0225",
    journal = "PoS",
    volume = "QNP2024",
    pages = "225",
    year = "2025"
}

@article{Zhao:2024ecc,
    author = "Zhao, Jiaxing and Aichelin, Joerg and Gossiaux, Pol Bernard and Ozvenchuk, Vitalii and Werner, Klaus",
    title = "{Heavy-flavor hadron production in relativistic heavy ion collisions at energies available at BNL RHIC and at the CERN LHC in the EPOS4HQ framework}",
    eprint = "2401.17096",
    archivePrefix = "arXiv",
    primaryClass = "hep-ph",
    doi = "10.1103/PhysRevC.110.024909",
    journal = "Phys. Rev. C",
    volume = "110",
    number = "2",
    pages = "024909",
    year = "2024"
}

@article{He:2025sfi,
    author = "He, Jialin and Peng, Xinye and Zhang, Xiaoming and Zheng, Liang",
    title = "{Studies of beauty hadron and non-prompt charm hadron production in pp collisions at $\sqrt{s}$=13 TeV within a transport model approach}",
    eprint = "2509.20291",
    archivePrefix = "arXiv",
    primaryClass = "hep-ph",
    month = "9",
    year = "2025"
}

@article{Lin:2001zk,
    author = "Lin, Zi-wei and Ko, C. M.",
    title = "{Partonic effects on the elliptic flow at RHIC}",
    eprint = "nucl-th/0108039",
    archivePrefix = "arXiv",
    doi = "10.1103/PhysRevC.65.034904",
    journal = "Phys. Rev. C",
    volume = "65",
    pages = "034904",
    year = "2002"
}

@article{ALICE:2023gjj,
    author = "Acharya, Shreyasi and others",
    collaboration = "ALICE",
    title = "{Measurement of non-prompt ${{\textrm{D}}^{0}}$-meson elliptic flow in Pb{\textendash}Pb collisions at $\sqrt{s_{\textrm{NN}}} = 5.02$~TeV}",
    eprint = "2307.14084",
    archivePrefix = "arXiv",
    primaryClass = "nucl-ex",
    reportNumber = "CERN-EP-2023-148",
    doi = "10.1140/epjc/s10052-023-12259-3",
    journal = "Eur. Phys. J. C",
    volume = "83",
    number = "12",
    pages = "1123",
    year = "2023"
}

@article{ALICE:2022tji,
    author = "Acharya, Shreyasi and others",
    collaboration = "ALICE",
    title = "{Measurement of beauty production via non-prompt D$^{0}$ mesons in Pb-Pb collisions at $ \sqrt{{\textrm{s}}_{\textrm{NN}}} $= 5.02 TeV}",
    eprint = "2202.00815",
    archivePrefix = "arXiv",
    primaryClass = "nucl-ex",
    reportNumber = "CERN-EP-2022-015",
    doi = "10.1007/JHEP12(2022)126",
    journal = "JHEP",
    volume = "12",
    pages = "126",
    year = "2022"
}

@article{Buncic:2015ari,
    author = "Buncic, P. and Krzewicki, M. and Vande Vyvre, P.",
    title = "{Technical Design Report for the Upgrade of the Online-Offline Computing System}",
    report = "\href{https://cds.cern.ch/record/2011297}{CERN-LHCC-2015-006, ALICE-TDR-019}",
    month = "4",
    year = "2015"
}

@article{Chen:2016XST,
    author = "Chen, Tianqi and Guestrin, Carlos",
     title = {{XGBoost}: A Scalable Tree Boosting System},
    booktitle = {Proceedings of the 22nd ACM SIGKDD International Conference on Knowledge Discovery and Data Mining},
    isbn = {978-1-4503-4232-2},
     eprint = "1603.02754",
    archivePrefix = "arXiv",
    primaryClass = "cs.LG",
    doi = "10.1145/2939672.2939785",
    month = "3",
    year = "2016"
}

@article{ALICETPC:2020ann,
    author = "Adolfsson, J. and others",
    collaboration = "ALICE TPC",
    title = "{The upgrade of the ALICE TPC with GEMs and continuous readout}",
    eprint = "2012.09518",
    archivePrefix = "arXiv",
    primaryClass = "physics.ins-det",
    doi = "10.1088/1748-0221/16/03/P03022",
    journal = "JINST",
    volume = "16",
    number = "03",
    pages = "P03022",
    year = "2021"
}

@article{ALICETOF:2025tof,
    author = "Abualrob, Ibrahim Jaser and others",
      collaboration = "ALICE",
      title         = "{Time resolution of the ALICE Time-Of-Flight detector with
                       the first Run 3 pp collisions at ${\sqrt{\textit{s}} =
                       13.6}$ TeV}",
      institution   = "CERN",
      archivePrefix = "arXiv",
      eprint        = "2511.10311",
      reportNumber  = "CERN-EP-2025-255",
      address       = "Geneva",
      year          = "2025"
}

@article{Lettrich:2752837,
      author        = "Lettrich, Michael",
      collaboration = "ALICE",
      title         = "{Fast Entropy Coding for ALICE Run 3}",
      archivePrefix = "arXiv",
      eprint        = "2102.09649",
      journal       = "PoS",
      volume        = "ICHEP2020",
      pages         = "913",
      year          = "2021",
      url           = "https://cds.cern.ch/record/2752837",
      doi           = "10.22323/1.390.0913",
}

@article{Bierlich:2022pfr,
    author = "Bierlich, Christian and others",
    title = "{A comprehensive guide to the physics and usage of PYTHIA 8.3}",
    eprint = "2203.11601",
    archivePrefix = "arXiv",
    primaryClass = "hep-ph",
    reportNumber = "LU-TP 22-16, MCNET-22-04, FERMILAB-PUB-22-227-SCD",
    doi = "10.21468/SciPostPhysCodeb.8",
    journal = "SciPost Phys. Codeb.",
    volume = "2022",
    pages = "8",
    year = "2022"
}

@article{Christiansen:2015yqa,
    author = "Christiansen, Jesper R. and Skands, Peter Z.",
    title = "{String Formation Beyond Leading Colour}",
    eprint = "1505.01681",
    archivePrefix = "arXiv",
    primaryClass = "hep-ph",
    reportNumber = "COEPP-MN-15-1, LU-TP-15-16, MCNET-15-09, COEPP-MN-15-1, LU-TP-15-16, MCNET-15-09",
    doi = "10.1007/JHEP08(2015)003",
    journal = "JHEP",
    volume = "08",
    pages = "003",
    year = "2015"
}

@article{GEANT4:2002zbu,
    author = "Agostinelli, S. and others",
    collaboration = "GEANT4",
    title = "{GEANT4 - A Simulation Toolkit}",
    reportNumber = "SLAC-PUB-9350, FERMILAB-PUB-03-339, CERN-IT-2002-003",
    doi = "10.1016/S0168-9002(03)01368-8",
    journal = "Nucl. Instrum. Meth. A",
    volume = "506",
    pages = "250--303",
    year = "2003"
}

@article{Trzaska:2017reu,
    author = "Trzaska, Wladyslaw Henryk",
    editor = "Badurek, G. and Bergauer, T. and Dragicevic, M. and Friedl, M. and Jeitler, M. and Krammer, M. and Schieck, J. and Schwanda, C.",
    collaboration = "ALICE",
    title = "{New Fast Interaction Trigger for ALICE}",
    doi = "10.1016/j.nima.2016.06.029",
    journal = "Nucl. Instrum. Meth. A",
    volume = "845",
    pages = "463--466",
    year = "2017"
}

@article{Hou:2019efy,
    author = "Hou, Tie-Jiun and others",
    title = "{New CTEQ global analysis of quantum chromodynamics with high-precision data from the LHC}",
    eprint = "1912.10053",
    archivePrefix = "arXiv",
    primaryClass = "hep-ph",
    reportNumber = "MSUHEP-19-025, PITT-PACC-1911, SMU-HEP-19-03",
    doi = "10.1103/PhysRevD.103.014013",
    journal = "Phys. Rev. D",
    volume = "103",
    number = "1",
    pages = "014013",
    year = "2021"
}

@article{ALICE:2022wpn,
    author = "Acharya, Shreyasi and others",
    collaboration = "ALICE",
    title = "{The ALICE experiment: a journey through QCD}",
    eprint = "2211.04384",
    archivePrefix = "arXiv",
    primaryClass = "nucl-ex",
    reportNumber = "CERN-EP-2022-227",
    doi = "10.1140/epjc/s10052-024-12935-y",
    journal = "Eur. Phys. J. C",
    volume = "84",
    number = "8",
    pages = "813",
    year = "2024"
}

\newpage
\appendix
\section{Comparison with existing measurements}
\label{app:cms_lhcb}
\begin{figure}[!htbp]
    \centering
    \includegraphics[width=0.8\linewidth]{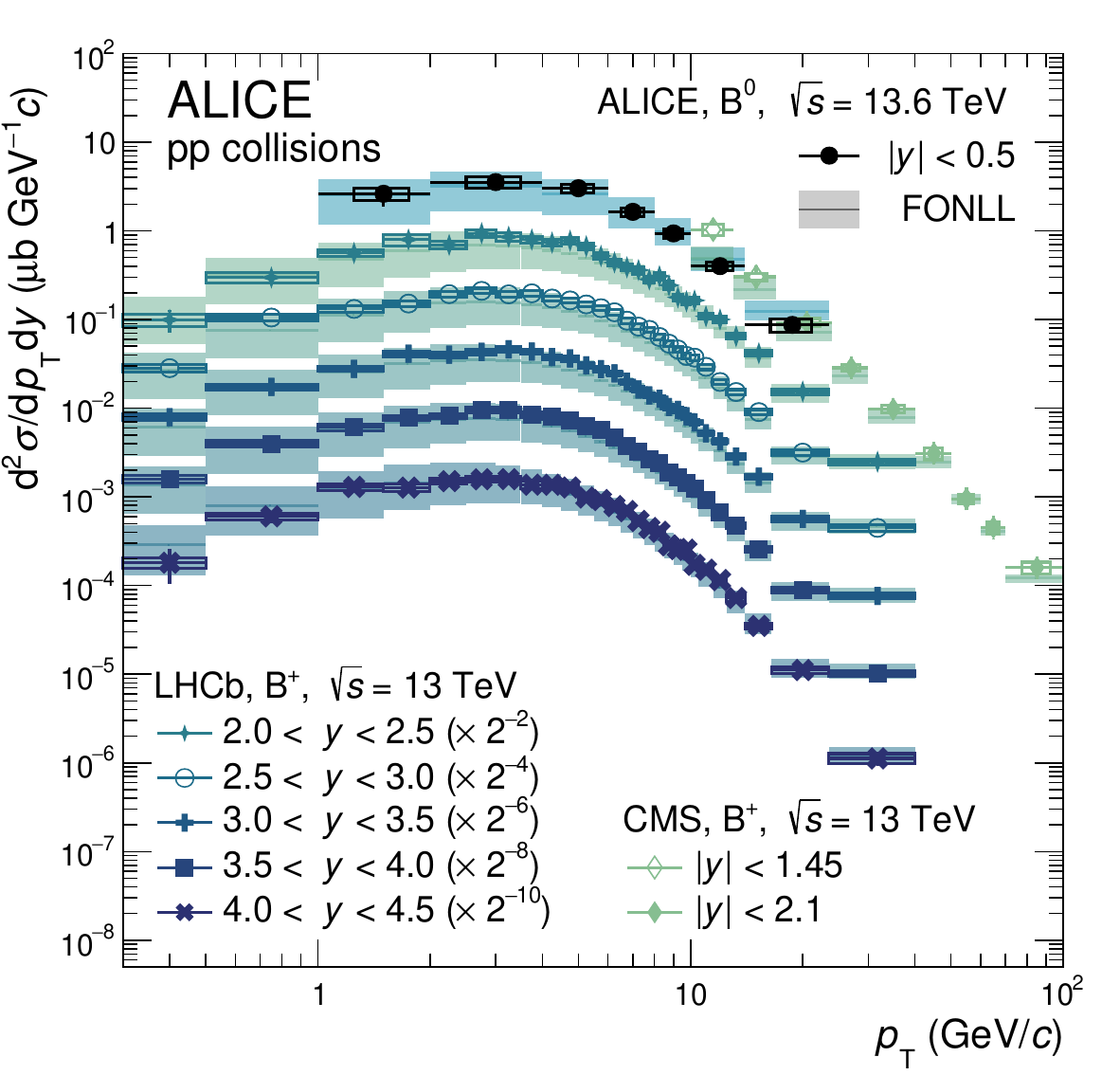}
    \caption{\pt-differential production cross section of \Bzero mesons at midrapidity ($|y|<0.5$) at the centre-of-mass energy of $\sqrts=13.6$~\tev compared with measurements of the production cross section of \Bplus mesons performed at $\sqrts=13$~\tev by the CMS Collaboration at midrapidity~\cite{CMS:2016plw} and by the LHCb Collaboration at forward rapidities~\cite{LHCb:2017vec}. The FONLL~\cite{ Cacciari:2012ny} calculations for the corresponding \pt and rapidity intervals are also shown as filled boxes.}
    \label{fig:cms_lhcb}
\end{figure}

The measurement reported in this article extends previous measurements performed at the LHC~\cite{ATLAS:2013cia,CMS:2011pdu,CMS:2016plw,CMS:2024vip,LHCb:2017vec} in both the \pt and rapidity ranges. Figure~\ref{fig:cms_lhcb} presents the \pt-differential production cross section of \Bzero mesons at midrapidity ($|y|<0.5$) at the centre-of-mass energy of $\sqrts=13.6$~\tev, and compares it with measurements of the production cross section of \Bplus mesons performed at $\sqrts=13$~\tev by the CMS Collaboration at midrapidity~\cite{CMS:2016plw} and by the LHCb Collaboration at forward rapidities~\cite{LHCb:2017vec}. The measurements are compared to FONLL~\cite{Cacciari:2012ny} calculations for the corresponding \pt and rapidity intervals.

\clearpage

%
%

\section{The ALICE Collaboration}
\label{app:collab}
\begin{flushleft} 
\small

D.A.H.~Abdallah\,\orcidlink{0000-0003-4768-2718}\,$^{\rm 134}$, 
I.J.~Abualrob\,\orcidlink{0009-0005-3519-5631}\,$^{\rm 112}$, 
S.~Acharya\,\orcidlink{0000-0002-9213-5329}\,$^{\rm 49}$, 
K.~Agarwal\,\orcidlink{0000-0001-5781-3393}\,$^{\rm II,}$$^{\rm 23}$, 
G.~Aglieri Rinella\,\orcidlink{0000-0002-9611-3696}\,$^{\rm 32}$, 
L.~Aglietta\,\orcidlink{0009-0003-0763-6802}\,$^{\rm 24}$, 
N.~Agrawal\,\orcidlink{0000-0003-0348-9836}\,$^{\rm 25}$, 
Z.~Ahammed\,\orcidlink{0000-0001-5241-7412}\,$^{\rm 132}$, 
S.~Ahmad\,\orcidlink{0000-0003-0497-5705}\,$^{\rm 15}$, 
I.~Ahuja\,\orcidlink{0000-0002-4417-1392}\,$^{\rm 36}$
% ...

\section*{Affiliation Notes}

$^{\rm I}$ Deceased\\
$^{\rm II}$ Also at: INFN Trieste\\
$^{\rm III}$ Also at: Fondazione Bruno Kessler (FBK), Trento, Italy\\
$^{\rm IV}$ Also at: Max-Planck-Institut fur Physik, Munich, Germany\\
$^{\rm V}$ Also at: Czech Technical University in Prague (CZ)\\
$^{\rm VI}$ Also at: Instituto de Fisica da Universidade de Sao Paulo\\
$^{\rm VII}$ Also at: Dipartimento DET del Politecnico di Torino, Turin, Italy\\
$^{\rm VIII}$ Also at: Department of Applied Physics, Aligarh Muslim University, Aligarh, India\\
$^{\rm IX}$ Also at: Institute of Theoretical Physics, University of Wroclaw, Poland\\
$^{\rm X}$ Also at: Facultad de Ciencias, Universidad Nacional Aut\'{o}noma de M\'{e}xico, Mexico City, Mexico\\

\section*{Collaboration Institutes}

$^{1}$ A.I. Alikhanyan National Science Laboratory (Yerevan Physics Institute) Foundation, Yerevan, Armenia\\
$^{2}$ AGH University of Krakow, Cracow, Poland\\
$^{3}$ Bogolyubov Institute for Theoretical Physics, National Academy of Sciences of Ukraine, Kyiv, Ukraine\\
$^{4}$ Bose Institute, Department of Physics  and Centre for Astroparticle Physics and Space Science (CAPSS), Kolkata, India\\
$^{5}$ California Polytechnic State University, San Luis Obispo, California, United States\\
$^{6}$ Central China Normal University, Wuhan, China\\
$^{7}$ Centro de Aplicaciones Tecnol\'{o}gicas y Desarrollo Nuclear (CEADEN), Havana, Cuba\\
$^{8}$ Centro de Investigaci\'{o}n y de Estudios Avanzados (CINVESTAV), Mexico City and M\'{e}rida, Mexico\\
$^{9}$ Chicago State University, Chicago, Illinois, United States\\
$^{10}$ China Nuclear Data Center, China Institute of Atomic Energy, Beijing, China\\
$^{11}$ China University of Geosciences, Wuhan, China\\
$^{12}$ Chungbuk National University, Cheongju, Republic of Korea\\
$^{13}$ Comenius University Bratislava, Faculty of Mathematics, Physics and Informatics, Bratislava, Slovak Republic\\
$^{14}$ Creighton University, Omaha, Nebraska, United States\\
$^{15}$ Department of Physics, Aligarh Muslim University, Aligarh, India\\
$^{16}$ Department of Physics, Pusan National University, Pusan, Republic of Korea\\
$^{17}$ Department of Physics, Sejong University, Seoul, Republic of Korea\\
$^{18}$ Department of Physics, University of California, Berkeley, California, United States\\
$^{19}$ Department of Physics, University of Oslo, Oslo, Norway\\
$^{20}$ Department of Physics and Technology, University of Bergen, Bergen, Norway\\
$^{21}$ Dipartimento di Fisica, Universit\`{a} di Pavia, Pavia, Italy\\
$^{22}$ Dipartimento di Fisica dell'Universit\`{a} and Sezione INFN, Cagliari, Italy\\
$^{23}$ Dipartimento di Fisica dell'Universit\`{a} and Sezione INFN, Trieste, Italy\\
$^{24}$ Dipartimento di Fisica dell'Universit\`{a} and Sezione INFN, Turin, Italy\\
$^{25}$ Dipartimento di Fisica e Astronomia dell'Universit\`{a} and Sezione INFN, Bologna, Italy\\
$^{26}$ Dipartimento di Fisica e Astronomia dell'Universit\`{a} and Sezione INFN, Catania, Italy\\
$^{27}$ Dipartimento di Fisica e Astronomia dell'Universit\`{a} and Sezione INFN, Padova, Italy\\
$^{28}$ Dipartimento di Fisica `E.R.~Caianiello' dell'Universit\`{a} and Gruppo Collegato INFN, Salerno, Italy\\
$^{29}$ Dipartimento DISAT del Politecnico and Sezione INFN, Turin, Italy\\
$^{30}$ Dipartimento di Scienze MIFT, Universit\`{a} di Messina, Messina, Italy\\
$^{31}$ Dipartimento Interateneo di Fisica `M.~Merlin' and Sezione INFN, Bari, Italy\\
$^{32}$ European Organization for Nuclear Research (CERN), Geneva, Switzerland\\
$^{33}$ Faculty of Electrical Engineering, Mechanical Engineering and Naval Architecture, University of Split, Split, Croatia\\
$^{34}$ Faculty of Nuclear Sciences and Physical Engineering, Czech Technical University in Prague, Prague, Czech Republic\\
$^{35}$ Faculty of Physics, Sofia University, Sofia, Bulgaria\\
$^{36}$ Faculty of Science, P.J.~\v{S}af\'{a}rik University, Ko\v{s}ice, Slovak Republic\\
$^{37}$ Faculty of Technology, Environmental and Social Sciences, Bergen, Norway\\
$^{38}$ Frankfurt Institute for Advanced Studies, Johann Wolfgang Goethe-Universit\"{a}t Frankfurt, Frankfurt, Germany\\
$^{39}$ Fudan University, Shanghai, China\\
$^{40}$ Gauhati University, Department of Physics, Guwahati, India\\
$^{41}$ Helmholtz-Institut f\"{u}r Strahlen- und Kernphysik, Rheinische Friedrich-Wilhelms-Universit\"{a}t Bonn, Bonn, Germany\\
$^{42}$ Helsinki Institute of Physics (HIP), Helsinki, Finland\\
$^{43}$ High Energy Physics Group,  Universidad Aut\'{o}noma de Puebla, Puebla, Mexico\\
$^{44}$ Horia Hulubei National Institute of Physics and Nuclear Engineering, Bucharest, Romania\\
$^{45}$ HUN-REN Wigner Research Centre for Physics, Budapest, Hungary\\
$^{46}$ Indian Institute of Technology Bombay (IIT), Mumbai, India\\
$^{47}$ Indian Institute of Technology Indore, Indore, India\\
$^{48}$ INFN, Laboratori Nazionali di Frascati, Frascati, Italy\\
$^{49}$ INFN, Sezione di Bari, Bari, Italy\\
$^{50}$ INFN, Sezione di Bologna, Bologna, Italy\\
$^{51}$ INFN, Sezione di Cagliari, Cagliari, Italy\\
$^{52}$ INFN, Sezione di Catania, Catania, Italy\\
$^{53}$ INFN, Sezione di Padova, Padova, Italy\\
$^{54}$ INFN, Sezione di Pavia, Pavia, Italy\\
$^{55}$ INFN, Sezione di Torino, Turin, Italy\\
$^{56}$ INFN, Sezione di Trieste, Trieste, Italy\\
$^{57}$ Inha University, Incheon, Republic of Korea\\
$^{58}$ Institute for Gravitational and Subatomic Physics (GRASP), Utrecht University/Nikhef, Utrecht, Netherlands\\
$^{59}$ Institute of Experimental Physics, Slovak Academy of Sciences, Ko\v{s}ice, Slovak Republic\\
$^{60}$ Institute of Physics, Homi Bhabha National Institute, Bhubaneswar, India\\
$^{61}$ Institute of Physics of the Czech Academy of Sciences, Prague, Czech Republic\\
$^{62}$ Institute of Space Science (ISS), Bucharest, Romania\\
$^{63}$ Institut f\"{u}r Kernphysik, Johann Wolfgang Goethe-Universit\"{a}t Frankfurt, Frankfurt, Germany\\
$^{64}$ Instituto de Ciencias Nucleares, Universidad Nacional Aut\'{o}noma de M\'{e}xico, Mexico City, Mexico\\
$^{65}$ Instituto de F\'{i}sica, Universidade Federal do Rio Grande do Sul (UFRGS), Porto Alegre, Brazil\\
$^{66}$ Instituto de F\'{\i}sica, Universidad Nacional Aut\'{o}noma de M\'{e}xico, Mexico City, Mexico\\
$^{67}$ iThemba LABS, National Research Foundation, Somerset West, South Africa\\
$^{68}$ Jeonbuk National University, Jeonju, Republic of Korea\\
$^{69}$ Korea Institute of Science and Technology Information, Daejeon, Republic of Korea\\
$^{70}$ Laboratoire de Physique Subatomique et de Cosmologie, Universit\'{e} Grenoble-Alpes, CNRS-IN2P3, Grenoble, France\\
$^{71}$ Lawrence Berkeley National Laboratory, Berkeley, California, United States\\
$^{72}$ Lund University Department of Physics, Division of Particle Physics, Lund, Sweden\\
$^{73}$ Marietta Blau Institute, Vienna, Austria\\
$^{74}$ Nagasaki Institute of Applied Science, Nagasaki, Japan\\
$^{75}$ Nara Women{'}s University (NWU), Nara, Japan\\
$^{76}$ National Centre for Nuclear Research, Warsaw, Poland\\
$^{77}$ National Institute of Science Education and Research, Homi Bhabha National Institute, Jatni, India\\
$^{78}$ National Nuclear Research Center, Baku, Azerbaijan\\
$^{79}$ National Research and Innovation Agency - BRIN, Jakarta, Indonesia\\
$^{80}$ Niels Bohr Institute, University of Copenhagen, Copenhagen, Denmark\\
$^{81}$ Nikhef, National institute for subatomic physics, Amsterdam, Netherlands\\
$^{82}$ Nuclear Physics Group, STFC Daresbury Laboratory, Daresbury, United Kingdom\\
$^{83}$ Nuclear Physics Institute of the Czech Academy of Sciences, Husinec-\v{R}e\v{z}, Czech Republic\\
$^{84}$ Oak Ridge National Laboratory, Oak Ridge, Tennessee, United States\\
$^{85}$ Ohio State University, Columbus, Ohio, United States\\
$^{86}$ Physics department, Faculty of science, University of Zagreb, Zagreb, Croatia\\
$^{87}$ Physics Department, Panjab University, Chandigarh, India\\
$^{88}$ Physics Department, University of Jammu, Jammu, India\\
$^{89}$ Physics Program and International Institute for Sustainability with Knotted Chiral Meta Matter (WPI-SKCM$^{2}$), Hiroshima University, Hiroshima, Japan\\
$^{90}$ Physikalisches Institut, Eberhard-Karls-Universit\"{a}t T\"{u}bingen, T\"{u}bingen, Germany\\
$^{91}$ Physikalisches Institut, Ruprecht-Karls-Universit\"{a}t Heidelberg, Heidelberg, Germany\\
$^{92}$ Physik Department, Technische Universit\"{a}t M\"{u}nchen, Munich, Germany\\
$^{93}$ Politecnico di Bari and Sezione INFN, Bari, Italy\\
$^{94}$ Research Division and ExtreMe Matter Institute EMMI, GSI Helmholtzzentrum f\"ur Schwerionenforschung GmbH, Darmstadt, Germany\\
$^{95}$ Saga University, Saga, Japan\\
$^{96}$ Saha Institute of Nuclear Physics, Homi Bhabha National Institute, Kolkata, India\\
$^{97}$ School of Physics and Astronomy, University of Birmingham, Birmingham, United Kingdom\\
$^{98}$ Secci\'{o}n F\'{\i}sica, Departamento de Ciencias, Pontificia Universidad Cat\'{o}lica del Per\'{u}, Lima, Peru\\
$^{99}$ SUBATECH, IMT Atlantique, Nantes Universit\'{e}, CNRS-IN2P3, Nantes, France\\
$^{100}$ Sungkyunkwan University, Suwon City, Republic of Korea\\
$^{101}$ Suranaree University of Technology, Nakhon Ratchasima, Thailand\\
$^{102}$ Technical University of Ko\v{s}ice, Ko\v{s}ice, Slovak Republic\\
$^{103}$ The Henryk Niewodniczanski Institute of Nuclear Physics, Polish Academy of Sciences, Cracow, Poland\\
$^{104}$ The University of Texas at Austin, Austin, Texas, United States\\
$^{105}$ Universidad Aut\'{o}noma de Sinaloa, Culiac\'{a}n, Mexico\\
$^{106}$ Universidade de S\~{a}o Paulo (USP), S\~{a}o Paulo, Brazil\\
$^{107}$ Universidade Estadual de Campinas (UNICAMP), Campinas, Brazil\\
$^{108}$ Universidade Federal do ABC, Santo Andre, Brazil\\
$^{109}$ Universitatea Nationala de Stiinta si Tehnologie Politehnica Bucuresti, Bucharest, Romania\\
$^{110}$ University of Cape Town, Cape Town, South Africa\\
$^{111}$ University of Derby, Derby, United Kingdom\\
$^{112}$ University of Houston, Houston, Texas, United States\\
$^{113}$ University of Jyv\"{a}skyl\"{a}, Jyv\"{a}skyl\"{a}, Finland\\
$^{114}$ University of Kansas, Lawrence, Kansas, United States\\
$^{115}$ University of Liverpool, Liverpool, United Kingdom\\
$^{116}$ University of Science and Technology of China, Hefei, China\\
$^{117}$ University of Silesia in Katowice, Katowice, Poland\\
$^{118}$ University of South-Eastern Norway, Kongsberg, Norway\\
$^{119}$ University of Tennessee, Knoxville, Tennessee, United States\\
$^{120}$ University of the Witwatersrand, Johannesburg, South Africa\\
$^{121}$ University of Tokyo, Tokyo, Japan\\
$^{122}$ University of Tsukuba, Tsukuba, Japan\\
$^{123}$ Universit\"{a}t M\"{u}nster, Institut f\"{u}r Kernphysik, M\"{u}nster, Germany\\
$^{124}$ Universit\'{e} Clermont Auvergne, CNRS/IN2P3, LPC, Clermont-Ferrand, France\\
$^{125}$ Universit\'{e} de Lyon, CNRS/IN2P3, Institut de Physique des 2 Infinis de Lyon, Lyon, France\\
$^{126}$ Universit\'{e} de Strasbourg, CNRS, IPHC UMR 7178, F-67000 Strasbourg, France, Strasbourg, France\\
$^{127}$ Universit\'{e} Paris-Saclay, Centre d'Etudes de Saclay (CEA), IRFU, D\'{e}partment de Physique Nucl\'{e}aire (DPhN), Saclay, France\\
$^{128}$ Universit\'{e}  Paris-Saclay, CNRS/IN2P3, IJCLab, Orsay, France\\
$^{129}$ Universit\`{a} degli Studi di Foggia, Foggia, Italy\\
$^{130}$ Universit\`{a} del Piemonte Orientale, Vercelli, Italy\\
$^{131}$ Universit\`{a} di Brescia, Brescia, Italy\\
$^{132}$ Variable Energy Cyclotron Centre, Homi Bhabha National Institute, Kolkata, India\\
$^{133}$ Warsaw University of Technology, Warsaw, Poland\\
$^{134}$ Wayne State University, Detroit, Michigan, United States\\
$^{135}$ Yale University, New Haven, Connecticut, United States\\
$^{136}$ Yildiz Technical University, Istanbul, Turkey\\
$^{137}$ Yonsei University, Seoul, Republic of Korea\\
$^{138}$ Affiliated with an institute formerly covered by a cooperation agreement with CERN\\
$^{139}$ Affiliated with an international laboratory covered by a cooperation agreement with CERN.\\

\end{flushleft} 
  
\end{document}